\documentclass[fleqn,usenatbib]{mnras}

\usepackage{newtxtext,newtxmath}

\usepackage[T1]{fontenc}
\usepackage{ae,aecompl}

\usepackage{graphicx}	
\usepackage{amsmath}	

\newcommand{\vect}[1]{\boldsymbol{#1}}

\title[AARTFAAC Cosmic Explorer]{The AARTFAAC Cosmic Explorer: observations of the 21-cm power spectrum in the EDGES absorption trough}

\author[B. K. Gehlot et al.]{B. K. Gehlot$^{1,2}$\thanks{E-mail: kbharatgehlot@gmail.com (BKG)}, F. G. Mertens$^{2,3}$, L. V. E. Koopmans$^{2}$\thanks{E-mail: koopmans@astro.rug.nl (LVEK)}, A. R. Offringa$^{2,4}$, 
\newauthor A. Shulevski$^{5,6}$, M. Mevius$^{4}$, M. A. Brentjens$^{4}$, M. Kuiack$^{5}$, V. N. Pandey$^{2,4}$,
\newauthor A. Rowlinson$^{4,5}$,  A. M. Sardarabadi$^{2}$, H. K. Vedantham$^{2,4}$, R. A. M. J. Wijers$^{5}$,
\newauthor S. Yatawatta$^{2,4}$ and S. Zaroubi$^{2,7}$
\\
$^{1}$School of Earth and Space Exploration, Arizona State University, Tempe, AZ, United States.\\
$^{2}$Kapteyn Astronomical Institute, University of Groningen, PO Box 800, 9700AV Groningen, The Netherlands.\\
$^{3}$LERMA, Observatoire de Paris, PSL Research University, CNRS, Sorbonne Universit\'e, F-75014 Paris, France.\\
$^{4}$ASTRON, Netherlands Institute for Radio Astronomy, Oude Hoogeveensedijk 4, 7991 PD, Dwingeloo, The Netherlands.\\
$^{5}$Anton Pannekoek Institute, University of Amsterdam, Postbus 94249 1090 GE, Amsterdam, The Netherlands.\\
$^{6}$Leiden Observatory, Leiden University, PO Box 9513, 2300 RA Leiden, The Netherlands.\\
$^{7}$Department of Natural Sciences, The Open University of Israel, 1 University Road, PO Box 808, Ra'anana 4353701, Israel.\\
}

\date{Accepted XXX. Received YYY; in original form ZZZ}

\pubyear{2020}

\begin{document}
\label{firstpage}
\pagerange{\pageref{firstpage}--\pageref{lastpage}}
\maketitle

\begin{abstract}
The 21-cm absorption feature reported by the EDGES collaboration is several times stronger than that predicted by traditional astrophysical models. If genuine, a deeper absorption may lead to stronger fluctuations on the 21-cm signal on degree scales (up to 1~Kelvin in rms), allowing these fluctuations to be detectable in nearly 50~times shorter integration times compared to previous predictions. We commenced the ``AARTFAAC Cosmic Explorer" (ACE) program, that employs the AARTFAAC wide-field imager, to measure or set limits on the power spectrum of the 21-cm fluctuations in the redshift range $z = 17.9-18.6$ ($\Delta\nu = 72.36-75.09$~MHz) corresponding to the deep part of the EDGES absorption feature. Here, we present first results from two LST bins: 23.5-23.75h and 23.5-23.75h, each with 2~h of data, recorded in `semi drift-scan' mode. We demonstrate the application of the new ACE data-processing pipeline (adapted from the LOFAR-EoR pipeline) on the AARTFAAC data. We observe that noise estimates from the channel and time-differenced Stokes~$V$ visibilities agree with each other. After 2~h of integration and subtraction of bright foregrounds, we obtain $2\sigma$ upper limits on the 21-cm power spectrum of $\Delta_{21}^2 < (8139~\textrm{mK})^2$ and $\Delta_{21}^2 < (8549~\textrm{mK})^2$ at $k = 0.144~h\,\textrm{cMpc}^{-1}$ for the two LST bins. Incoherently averaging the noise bias-corrected power spectra for the two LST bins yields an upper limit of $\Delta_{21}^2 < (7388~\textrm{mK})^2$ at $k = 0.144~h\,\textrm{cMpc}^{-1}$. These are the deepest upper limits thus far at these redshifts.
\end{abstract}


\section{Introduction}

Observations of the redshifted 21-cm signal of neutral hydrogen from the Cosmic Dawn and Epoch of Reionization hold the potential to revolutionise our understanding of how these first stars and galaxies formed and the nature of their ionising radiation \citep{madau1997,shaver1999,furlanetto2006,pritchard2012,mesinger2011,zaroubi2013}. During the Cosmic Dawn (CD) ($12\lesssim z \lesssim 30$), the first luminous objects formed in the dark and neutral Universe \citep{pritchard2007}. X-ray and ultraviolet radiation from these first stars heated and ionized neutral hydrogen (HI) in the surrounding Inter-Galactic Medium (IGM) during the Epoch of Reionization (EoR). This process continued (spanning the redshift range $6\lesssim z \lesssim 12$) until hydrogen in the IGM became fully ionized \citep{madau1997}.

In recent years, a large number of observational efforts got underway to observe this faint 21-cm signal from the CD and EoR. Radio interferometers such as the Giant Meterwave Radio Telescope (GMRT; \citealt{paciga2011}), the LOw Frequency ARray (LOFAR; \citealt{vanhaarlem2013}), the Murchison Widefield Array (MWA; \citealt{tingay2013,bowman2013}), the Precision Array for Probing the Epoch of Reionization (PAPER; \citealt{parsons2010}) as well as the next-generation instruments such as the Hydrogen Epoch of Reionization Array (HERA; \citealt{deboer2017}), the Long Wavelength Array (LWA; \citealt{greenhill2012}), the New Extension in Nan\c cay Upgrading loFAR (NENUFAR; \citealt{zarka2012}), and the upcoming Square Kilometre Array (SKA; \citealt{mellema2013,koopmans2015}) are working towards measuring the spatial brightness temperature fluctuations in the high-redshift cosmological 21-cm signal. 

In parallel, single-element radiometers such as the Experiment to Detect the Global Epoch of Reionization Signature (EDGES; \citealt{bowman2018}), the Large-aperture Experiment to Detect the Dark Ages (LEDA; \citealt{bernardi2016}), the Shaped Antenna measurement of the background RAdio Spectrum 2 (SARAS 2; \citealt{singh2017}), the Sonda Cosmol\'{o}gica de las Islas para la Detecci\'{o}n de Hidr\'{o}geno Neutro (SCI-HI; \citealt{voytek2014}), the Probing Radio Intensity at high $z$ from Marion (PRIZM; \citealt{philip2019}), and the Netherlands-China Low frequency Explorer\footnote{\url{https://www.ru.nl/astrophysics/research/radboud-radio-lab-0/projects/netherlands-china-low-frequency-explorer ncle/}}$^,$\footnote{\url{https://www.astron.nl/r-d-laboratory/ncle/netherlands-china-low-frequency-explorer-ncle}} (NCLE) are seeking to measure the global sky-averaged 21-cm signal as a function of redshift.

In 2018, a deep spectral feature centred at 78 MHz was reported by the EDGES collaboration \citep{bowman2018}. The feature was presented as the long sought-after 21-cm absorption feature seen against the CMB during the CD at $z\sim17$. The location of this putative absorption trough is consistent with redshift predictions from theoretical models and simulations of the Cosmic Dawn \citep{furlanetto2006,pritchard2010,mesinger2013,cohen2017}. However, the depth of the feature is $\Delta T_{21}\sim 0.5$~K ($99\%$ confidence level), which is $2-3$~times stronger and considerably wider ($\Delta \nu \sim 19$~MHz) than that predicted by the most optimistic astrophysical models (e.g. \citealt{pritchard2010,fialkov2014,fialkov2016,cohen2017}). Moreover, the observed feature is flat-bottomed instead of a smooth Gaussian-like shape. Several ``exotic" theoretical models have already been proposed which might explain the depth of the feature, such as a considerably colder IGM due to interaction between baryons and dark matter particles causing a lower spin-temperature and therefore a deeper absorption feature (e.g.  \citealt{barkana2018,fialkov2018}), or a stronger radiation background against which the absorption is taking place (e.g. \citealt{feng2018,ewall-wice2018,dowell2018,fialkov2019}). Although the 21-cm signal is expected to be stronger at these redshifts, the foreground emission is several times brighter at these frequencies compared to EoR 21-cm signal observations at 150~MHz \citep{bernardi2009,bernardi2010}. Moreover, ionospheric effects are amplified at lower frequencies \citep{gehlot2018a,degasperin2018}, rendering the measurement of the signal equally (or even more) challenging than in EoR experiments. As of now, \cite{ewall-wice2016} reported a systematics-limited power spectrum upper limit of $\Delta_{21}^2 < (10^4~\text{mK})^2$ on co-moving scales $k\lesssim 0.5~h\,\text{cMpc}^{-1}$ (in 3~h of integration) on the 21-cm signal brightness temperature in the redshift range $12\lesssim z \lesssim 18$ using MWA. This overlaps with the low redshift edge of the 21-cm absorption feature \citep{bowman2018}. \cite{gehlot2019} provided a $2\sigma$ upper limit of $\Delta_{21}^2 < (1.4\times 10^4~\text{mK})^2$ on the 21-cm signal power spectrum at $k = 0.038~h\,\text{cMpc}^{-1}$ (in 14~h of integration) using the LOFAR-Low Band Antenna (LBA) system in the redshift range $19.8\lesssim z \lesssim 25.2$, which corresponds to the high redshift edge of the absorption feature. More recently, \cite{eastwood2019} used OVRO-LWA observations to report a $2\sigma$ upper limit of $\Delta_{21}^2 < (10^4~\text{mK})^2$ at $k \approx 0.1~h\,\text{cMpc}^{-1}$ (in 28~h of integration) at redshift $z\approx 18.4$. 

Although concerns have been raised about the validity of the detection of the absorption feature in terms of foreground modelling and instrumental effects \citep{hills2018,bradley2019}, if the detection is confirmed, the strength of the 21-cm absorption feature can also cause a significant increase in the 21-cm brightness temperature fluctuations in the redshift range $z = 17-19$ \citep{barkana2018,fialkov2018}. This redshift range corresponds to the deepest part of the absorption profile. It enables detection of the 21-cm signal brightness temperature fluctuations on degree angular scales in this redshift range within a much shorter integration time ($\sim50$~times shorter) compared to what was previously expected. 

\begin{figure*}
    \centering
    \includegraphics[width=\textwidth]{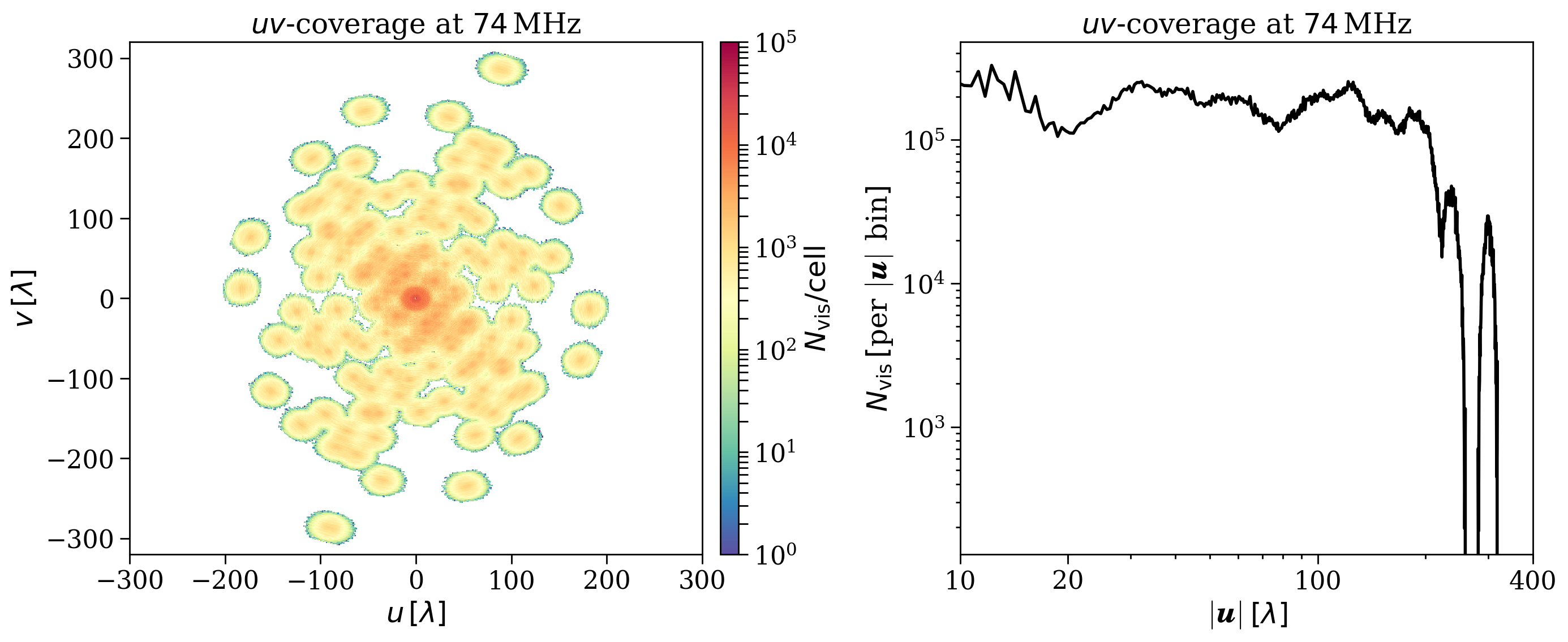}
    \caption{The left panel shows the $uv$-coverage of \texttt{A12} mode (12~station configuration) of the LOFAR-AARTFAAC LBA array at 74~MHz for 15~min of synthesis. The right panel shows the radial profile of the uv-coverage i.e. $N_{\text{vis}}$ per baseline bin ($|\vect{u}| =\sqrt{u^2+v^2} $) of width $0.5\lambda$.} 
    \label{fig:layout}
\end{figure*} 

Motivated by this, we have commenced a large scale program called ``AARTFAAC Cosmic Explorer" (ACE) to measure or limit the power spectrum of the brightness temperature fluctuations of the 21-cm signal from $z\sim 18$ using the LOFAR Amsterdam-ASTRON Radio Transients Facility And Analysis Centre (AARTFAAC) wide-field imager \citep{prasad2016}. AARTFAAC correlates up to 576 individual receiver elements (LBA dipoles or High Band Antenna tiles) in the core of LOFAR, thereby providing a wider field of view and increased sensitivity on large angular-scales compared to regular LOFAR observations. The redshift range targeted by ACE is $z=17.9-18.6$ ($72.36-75.09$~MHz frequency range) that corresponds to the deep part of the EDGES absorption feature. The ACE programme\footnote{LOFAR proposal ID: LT10\_006. Investigators: Gehlot and Koopmans} has collected about 500~h deep integration data of a large part of the northern sky to measure the power spectrum. In this work, we present first power spectrum results in two LST bins each with 2~h of data and successfully demonstrate the end-to-end application of the new ACE data-processing pipeline which is adapted from the LOFAR-EoR data processing pipeline to AARTFAAC data. Readers may refer to \cite{patil2017,gehlot2019,mertens2020} for an overview of the LOFAR-EoR data processing pipeline, and a description of HBA and LBA data processing.  

The paper is organised as follows: Section~\ref{sec:obs_n_prep} briefly describes the AARTFAAC wide-field imager, the observation setup of ACE observations, and the basic preprocessing steps for the raw data, e.g. flagging and averaging. The calibration and imaging strategy for the ACE data is described in Section~\ref{sec:calibration-imaging}. In Section~\ref{sec:Noise-stat}, we estimate and discuss the noise in ACE data and method of combining multiple ACE observations. In Section~\ref{sec:GPR_fg_removal}, we describe the Gaussian Process Regression foreground removal technique and power spectrum estimation methodology. We discuss results from the analysis Section~\ref{sec:multi_night_results}. Finally, we summarise the work and discuss future outlook in section \ref{sec:summary}. We use $\Lambda$CDM cosmology throughout the analyses with cosmological parameters consistent with Planck \citep{planck2015cosmo}.

\section{Observations and preprocessing}\label{sec:obs_n_prep}

We used the LOFAR-AARTFAAC wide-field imager to observe the northern sky in the frequency range $72.4-75$~MHz. The sky was observed in 'semi drift-scan' mode\footnote{AARTFAAC-LBA (unlike HBA) does not beam-form to track; it only points to the instantaneous zenith direction per integration time in drift-scan mode. During preprocessing, the drift-scan data for a long observational run are split into 15~min observation blocks and rephased. The phase centre for each block follows a constant declination point that passes through the zenith half-way during the 15~min run. We refer to this strategy as the 'semi drift-scan' mode.}, and the observed snapshot data was processed using a tweaked version of the LOFAR-EoR data processing pipeline \citep{gehlot2018a,gehlot2019,mertens2020}. The observational setup and the preprocessing steps are briefly described in following subsections.

\begin{table*}
	\centering
	\caption{Observational and correlator setting details.}
	\label{tab:obs_details}
	\begin{tabular}{ll} 
		\hline		
		\textbf{Parameter} & \textbf{value} \\	
		\hline
		Telescope & LOFAR AARTFAAC \\
		Observation cycle and ID & Cycle 10, LT10\_006\\ 		
		Antenna configuration & \texttt{A12} \\
		Number of receivers & 576 (LBA dipoles) \\
		Sidereal bins (h) & 23.5-23.75\,h and 23.75-24.00\,h \\
        Number of observation blocks (per LST-bin) & 8   \\
		Phase Center  & Bin 1: RA 23h37m30s, Dec +52d38m00s \\
		              & Bin 2: RA 23h52m30s, Dec +52d38m00s \\ 
        Minimum frequency & 72.36 MHz \\
        Maximum frequency & 75.09 MHz \\
		Target bandwidth & 2.73 MHz \\
        Outrigger sub-bands & 68.36 MHz and 78.90 MHz\\
		Primary Beam FWHM & $120^{\circ}$ at 74 MHz \\
		Field of View & 11000 $\text{deg}^2$ at 74 MHz\\	
		Polarization & Linear X-Y   \\
		Time, frequency resolution: \\
		\quad Raw Data & 1 s, 65.1 kHz      \\
		\quad After flagging and averaging & 4 s, 65.1 kHz \\
		\hline
	\end{tabular}
\end{table*}

\subsection{The LOFAR AARTFAAC wide-field imager}\label{subsec:ACE_imager}

The Amsterdam-ASTRON Radio Transient Facility and Analysis Centre (AARTFAAC) is a LOFAR based all-sky radio transient monitor \citep{prasad2016,kuiack2019}. It piggybacks on ongoing LOFAR observations and taps the digital signal streams of individual antenna elements from six or twelve core stations depending on the requirements. AARTFAAC operates in two modes viz. \texttt{A6} where the six innermost stations (also called the ``superterp") of the LOFAR-core are used, and \texttt{A12} that employs twelve innermost stations of the LOFAR-core. The \texttt{A6} mode consists of 288~dual-polarization receivers (e.g. Low Band Antenna (LBA) dipoles or High Band Antenna (HBA) tiles) within a 300~m diameter circle, and the \texttt{A12} mode consists of 576 such receivers spread across 1.2~km \citep{vanhaarlem2013}. Figure \ref{fig:layout} shows the 15~min synthesis $uv$-coverage of the \texttt{A12} mode at 74~MHz and the radial profile ($N_{\text{vis}}$ per baseline bin of $d\vect{u} = 0.5\lambda$) of the $uv$-coverage. The latter is relatively flat between $\vect{u} = 10-100\lambda$ except for a dip around $\vect{u}\sim 20\lambda,\,70\lambda$ which is due to slightly patchy LBA-dipole layout in the ``superterp" and the transition to non-superterp baselines. The array is co-planar at centimetre level within $0-70\lambda$, which is beneficial for wide-field imaging. In addition to this, the baselines up to 1.2~km support intermediate resolution imaging which helps to improve calibration and better captures compact structure in the sky. Each of the inner twelve LOFAR core stations consists of 96~LBA dipoles \footnote{LBA dipoles are dual-polarization (X-Y) dipoles optimised to operate between 30-80~MHz} (only 48 out of 96 dipoles can be used at a time) and 48 HBA tiles \footnote{HBA tiles consist of 16 dual-polarization dipoles arranged in a $4\times 4$ grid, which are analogue beam-formed to produce a single tile beam. HBA tiles are optimised to operate between 110-240~MHz.}. At a given time, AARTFAAC can only observe in either LBA or HBA mode depending upon the ongoing LOFAR observation. The digitised signal from the corresponding receiver elements is tapped and transported to the AARTFAAC correlator (located at the Centre for Information Technology (CIT)\footnote{\url{https://www.rug.nl/society-business/centre-for-information-technology/}} in Groningen, Netherlands) prior to beam-forming. Due to network limitations, only 16~sub-bands can be correlated using the 16-bit mode. Each sub-band is 195.3~kHz wide and consists of up to 64~channels providing a maximum frequency resolution of 3~kHz, with currently maximum instantaneous system bandwidth of 3.1~MHz. The correlator subsystem is a GPU based and produces correlations (XX, XY, YX, YY) for all dipoles pairs for every frequency channel with 1~s integration. The correlator has 1152 input streams with 576 signal streams per polarization. The output correlations can either be dumped as raw correlations on storage disks on the AARTFAAC storage/compute cluster or can be routed to the AARTFAAC real-time calibration and imaging pipeline for transient detection. AARTFAAC can only observe in drift-scan mode. However, phase-tracking can be applied to raw data during or after preprocessing. The raw data from the AARTFAAC storage/compute cluster can be streamed via a fast network (1~Gbit/s) to the LOFAR-EoR processing cluster `Dawn' at the CIT. The raw data can be converted to standard Measurement Set (MS) format using a custom software package \textsc{aartfaac2ms}\footnote{\url{https://github.com/aroffringa/aartfaac2ms}} \citep{offringa2015} which can also apply offline phase tracking. Readers may refer to \cite{prasad2016} for further information about AARTFAAC system design and capabilities, and \cite{vanhaarlem2013} for observing capabilities of LOFAR. 

\subsection{ACE observational setup and status}\label{subsec:observations}

We use the \texttt{A12} mode of AARTFAAC to observe the Northern sky in the drift-scan mode with the mean phase centre at the zenith. We use 14~contiguous sub-bands (a total of 2.73~MHz bandwidth) to observe the $72.36 - 75.09$~MHz, targeting the redshift range $z = 17.9 - 18.6$. We place the two remaining sub-bands $\sim 4$~MHz away from the targeted band centre on either side of the band, to aid in assessing the wide-band systematics and calibration quality as well as help foreground modelling and subtraction. We choose three channels per sub-band (with 65.1~kHz resolution) and 1~s correlator integration. High spectral and time resolution provides improved RFI excision and a better handle on delay/frequency transform (discussed in section \ref{subsec:power-spectra}). The ACE observing campaign concluded after observing around 500\,h of northern sky in drift-scan mode during 3~LOFAR observing cycles. Most observations span night time LSTs with a typical span of $4-12$~h per observation. 

For this pilot analysis, we select two LST bins, viz 23.5-23.75\,h (LST:23.5\,h or LST-bin~1 hereafter) and 23.75-24.00\,h (LST:23.75\,h or LST-bin~2 hereafter) with 8~observation blocks of 15~min each taken from different nights recorded during first cycle observations. This corresponds to the total integration time of 2~h per LST-bin. Table \ref{tab:obs_details} summarizes the observational and correlator setting details.

\begin{figure*}
\centering
\includegraphics[width=1\textwidth]{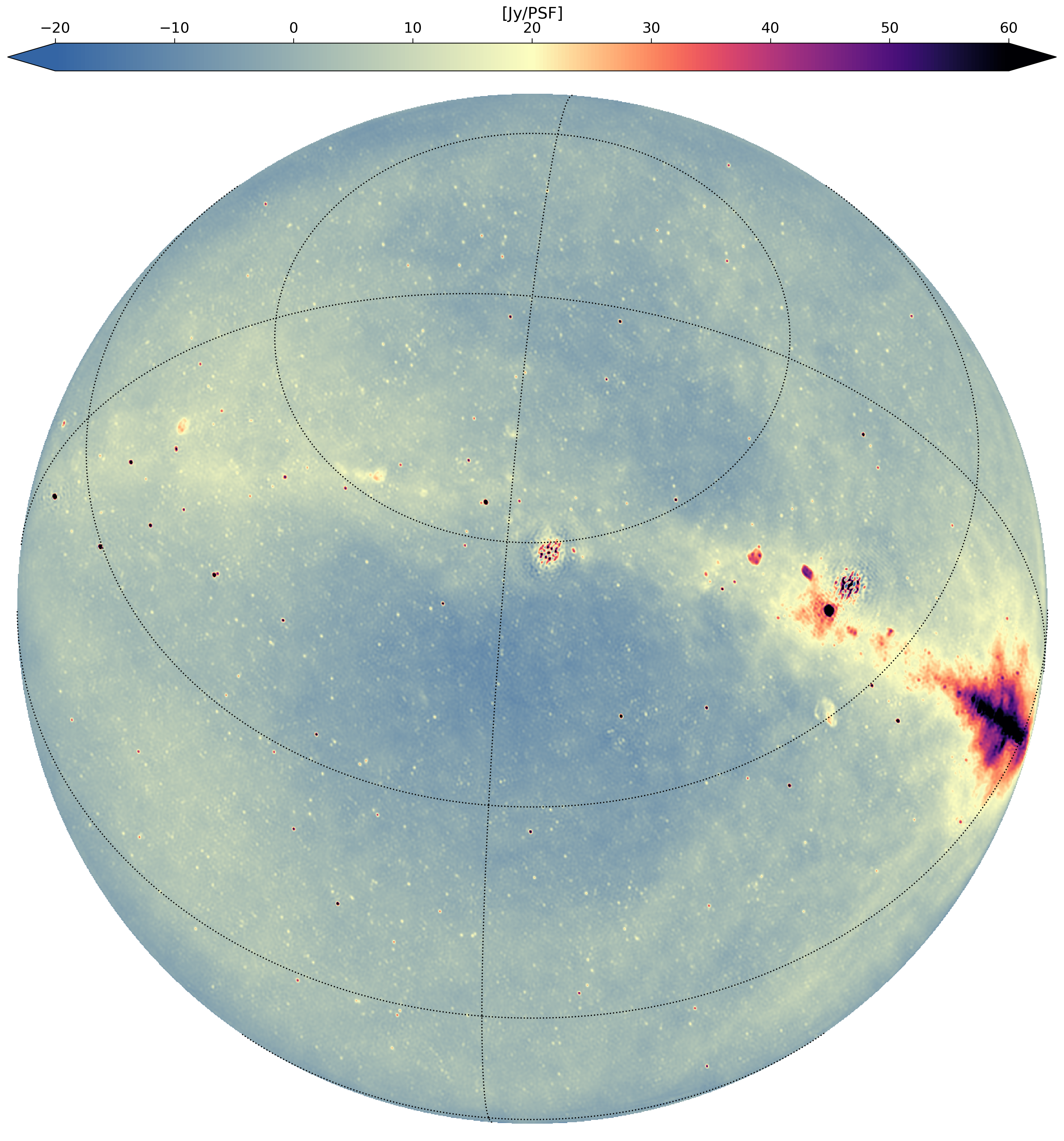}
    \caption{Intermediate resolution Stokes~$I$ continuum image ($72.4-75.1$~MHz, single night, cleaned) of LST:23.5h bin used in the analysis. The image is shown in orthographic projection, where dotted curves represent the parallels and meridians corresponding to DEC ($30^\circ$ separation) and RA, respectively.} 
\label{fig:mosaic}
\end{figure*} 

\subsection{Data preprocessing}\label{subsec:preprocessing}

The first step of data processing is to apply tracking to the drift-scan observations. For instruments with much wider Field of Views (FoV; FWHM$\sim120^{\circ}$ in our case), phase referencing to a single stationary point in the sky during long observations limits the portion of the sky which is visible. This is not an optimal strategy for long-duration observations. Therefore, instead of fixing the phase reference to a single stationary point for the entire observation, we choose to re-phase every 15~min observation block. The phase centre for each observation block is a constant declination point (on a great circle through zenith) which passes through zenith mid-way during the 15~min observation. We refer these re-phased 15~min observation blocks as `time-slices' throughout the paper.

The next step is RFI-excision, which is performed on the highest resolution data to minimize information loss. We use \textsc{aoflagger} \citep{offringa2010,offringa2012} to perform RFI excision on raw data and also flag all visibilities that include non-working LBA dipoles ($\sim 6-7\%$). The remaining data is averaged to a resolution of 4~s and 65.1~kHz and subsequently divided into 15~min time-slices for every individual phase centre. Each time-slice is separately written into Measurement Set (MS) format, which are stored permanently on the LOFAR-EoR processing cluster. The data volume in MS format is around 150~GB for 15~min time-slices, and $\sim1.2$~TB for 2~h worth of data, respectively. The \textsc{aartfaac2ms} package performs the re-phasing and flagging tasks and returns the phased and flagged data in MS format. The dipoles within a station share the same electronic cabinets, such that intra-station baselines may be affected by mutual-coupling/cross-talk effects. Therefore all intra-station baselines ($|\vect{b}| \lesssim 80\,\text{m}$) are also flagged post MS conversion.
The fraction of visibilities flagged by \textsc{aoflagger} (excluding non-working dipoles), at this stage, varies between $2-2.5\%$ for different time-slices in the two LST bins.

\section{Calibration and Imaging} \label{sec:calibration-imaging}

Visibilities measured by AARTFAAC are corrupted by the errors caused due to instrumental imperfections such as complex receiver gain, primary-beam and global band-pass, as well as environmental effects, for example, due to the ionosphere. Calibration of AARTFAAC refers to the estimation of these errors and correcting the observed visibilities to obtain a reliable estimate of the true sky visibilities. The errors that corrupt the visibilities can be classified into two broad categories: Direction Independent (DI) errors and Direction Dependent (DD) errors. DI errors are independent of the direction of the incoming signal from the sky and comprise of complex receiver gain and frequency band-pass, as well as a global ionospheric phase. On the contrary, DD errors change with sky direction, e.g. as a result of the antenna beam pattern, ionospheric phase fluctuations, and Faraday rotation \citep{hamaker1996a,hamaker1996b,smirnov2011a,smirnov2011b}. 

\subsection{Direction Independent calibration}\label{subsec:DIcal}
Direction Independent (DI) Calibration involves estimation of complex gains (full-Jones) per dipole, per time and frequency interval (represented by a complex $2\times2$ Jones matrix for two linear polarizations). We use \textsc{dppp}\footnote{\url{https://www.astron.nl/lofarwiki/doku.php?id=public:user_software:documentation:ndppp}} to calibrate the raw visibilities and subsequently apply the gain solutions obtained in the calibration to the visibilities. Unlike \textsc{sagecal-co} \citep{yatawatta2015,yatawatta2016,yatawatta2017} that we used previously in \cite{gehlot2019} for LBA-beam-formed data, \textsc{dppp} employs the primary beam model for individual LBA dipoles. We use Cas\,A and Cyg\,A (the two brightest sources in the northern sky) to calibrate the visibilities. Their sky-model consists of 14~components (9~components for Cas\,A and 5~components for Cyg\,A), i.e. Delta functions and Gaussians. The models of these sources were obtained using LOFAR-LBA observations and the source fluxes in the model, within a few per cent, are consistent with the Very Large Array (VLA) observations at 74~MHz \citep{cohen2007,kassim2007}. We use a power-law with a spectral index of $-0.8$ to represent the source spectra. We choose a calibration solution-time interval of 16~s for each 65.1~kHz channel to account for DI (or beam-averaged) instrumental and slower ionospheric effects while maintaining a reasonable signal-to-noise ratio ($\sim 30$) over the calibration interval. During calibration, we exclude the baselines $|\vect{u}| < 20\lambda$ in order to avoid the large-scale diffuse Galactic emission biasing the calibration solutions. We apply a LOFAR-LBA dipole beam model\footnote{Current LOFAR-LBA dipole beam models are based on Electro-Magnetic (EM) simulations of the LOFAR-LBA dipoles (private communication with LOFAR Radio Observatory).} during the model prediction step to adjust the flux scale. Absolute flux scale can be obtained by applying the beam model before the imaging step. Although the current sky-model is somewhat limited in terms of the number of sources, it represents most of the flux on the baselines used for later analyses. We are working on developing a more accurate sky-model for calibration, which will include compact sources above the confusion limit and multi-scale diffuse emission for robust calibration of AARTFAAC.

\begin{figure*}
\centering
\includegraphics[width=1\textwidth]{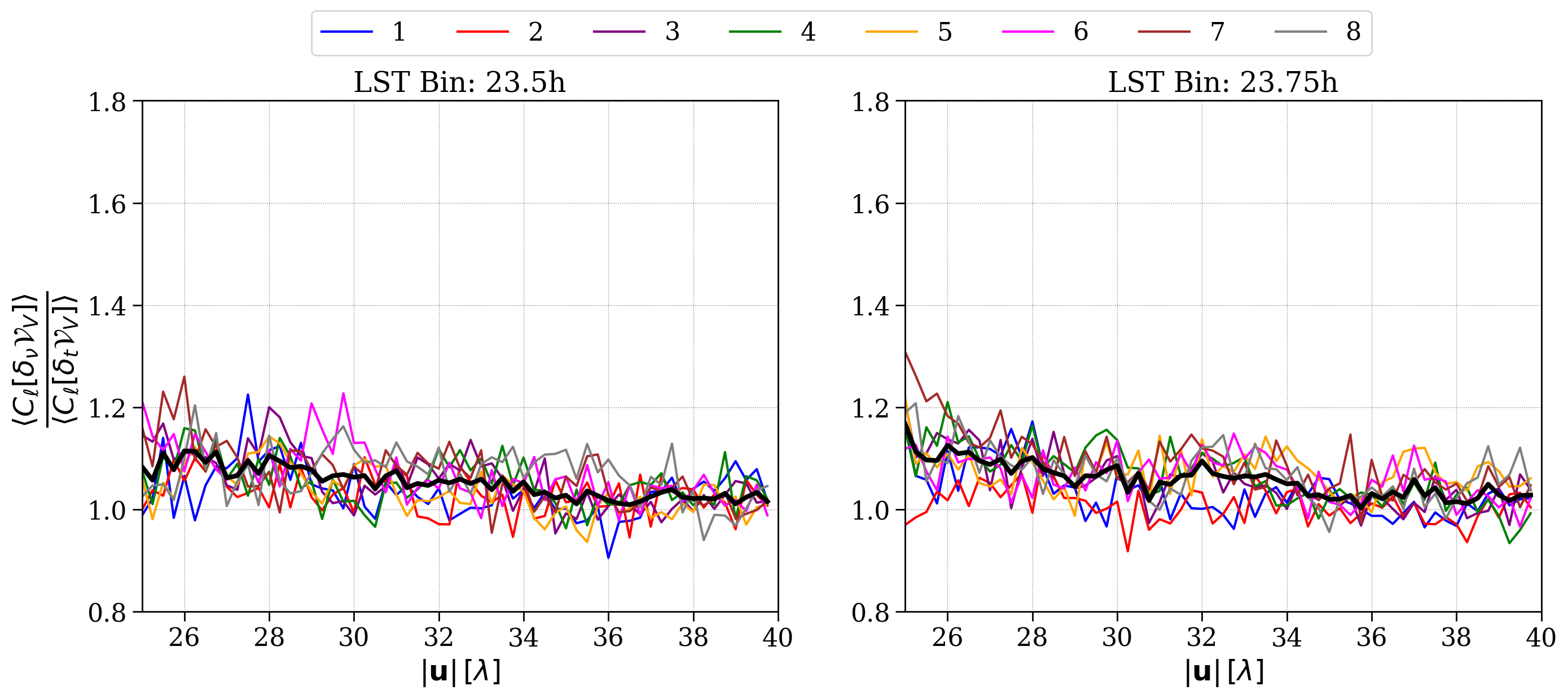}
    \caption{Ratio of frequency averaged angular power spectra of frequency and time difference visibilities. Left panel shows $\langle C_{\ell}[\delta_{\nu} \mathcal{V}_{V}] \rangle_t / \langle C_{\ell}[\delta_{t} \mathcal{V}_{V}]\rangle_{\nu}$ for different nights in LST:23.5\,h bin. Right panel is same as the left panel but for LST:23.75\,h bin. The black curve shows the mean of the ratios for 8 nights.} 
\label{fig:Noise-PS-ratio}
\end{figure*} 

\subsection{Direction Dependent calibration}\label{subsec:DDcal}

The two brightest sources Cas\,A and Cyg\,A dominate the visibilities and superpose significant PSF side-lobes over the field. It is crucial to subtract these sources to reduce the confusion due to these side-lobes. We use \textsc{dppp} (\texttt{DDEcal}) to subtract these sources using Direction Dependent (DD) calibration. We use a calibration solution interval of 16~s and 65.1~kHz, respectively. The calibration is constrained in the frequency direction and enforces frequency smoothness at 2~MHz level. This is somewhat similar to consensus optimisation in \textsc{sagecal-co} \citep{yatawatta2016,yatawatta2017}, which also enforces frequency smoothness of gain solutions. We exclude baselines $|\vect{u}| < 60\lambda$ in this step to reduce bias due to source subtraction on smaller baselines used for analyses in later sections. The directional gain solutions obtained towards Cas\,A and Cyg\,A are used to subtract them. 
 
After this, we perform another flagging step where we flag dipoles with relatively high visibility variance in time and frequency. We use \textsc{aoquality} and \textsc{aoqplot} (bundled with \textsc{aoflagger}) to generate quality statistics and flag dipoles with $>5$~ times the variance compared to the average. Subsequently, we run \textsc{aoflagger} and SSINS (Sky Subtracted Incoherent Noise Spectra; see \citealt{wilensky2019} for more details) based flagger to flag visibilities with bad (non-converged) solutions and corrupted by low-level RFI, respectively. After this intermediate flagging step, we re-perform the DI, and DD calibration steps in sections \ref{subsec:DIcal} and \ref{subsec:DDcal}, respectively. Finally, another instance of \textsc{aoflagger} is run to remove any remaining visibilities with bad/non-converged calibration solutions. After two rounds of flagging, the final fraction of flagged visibilities in the $20-50\lambda$ baseline range amounts to $4.5-7\%$ for different nights in the LST:23.5h bin, and $4-6\%$ in the LST:23.75h bin (except for two nights in the second bin, where flagged visibilities fractions are around $9\%$ and $14\%$, respectively).

\subsection{Imaging}\label{subsec:Imaging}

The visibilities after DI, DD-calibration and iterative flagging are imaged with \textsc{wsclean} package \citep{offringa2014,offringa2017}, a wide-field interferometric imaging software that uses the $w$-stacking algorithm. We use a `Kaiser-Bessel' kernel \citep{kaiser1980}, which is an approximation of the Prolate Spheroidal Wave Function \citep{jackson1991}, for gridding with a kernel-size of 31~pixels with an oversampling factor of $1.6\times10^4$ and a padding factor of~1.5 to avoid any artefacts due to gridding. Readers may refer to \cite{offringa2019} for a detailed analysis of convolutional gridding artefacts, their impact on 21-cm power spectra and methods to mitigate these artefacts. We use the $20-60\lambda$ baseline range with `natural' weighting scheme to produce Stokes~$I$, $V$ and PSF images for all channels and time-slices over the full visible sky for further analysis. The image-cubes produced by \textsc{wsclean} are converted to gridded visibilities in brightness temperature units of Kelvin. The cubes are trimmed to $120\deg$ size using a `Hann'\footnote{The `Hann' window is defined as: \\ $W(n) = 0.5 -0.5\cos \left[\dfrac{2\upi n}{(M-1)} \right]$, where $0\leq n\leq M-1$.} spatial taper (see e.g. \citealt{blackman1958}) and $25-40\lambda$ baseline range for further analyses. These gridded visibility cubes ($\mathcal{V}(u,v,\nu)$), number of visibilities per $uv$-cell ($N(u,v,\nu)$) and other related metadata are stored in HDF5 data format\footnote{\url{https://www.hdfgroup.org/solutions/hdf5}}. Figure \ref{fig:mosaic} shows a higher resolution deep-cleaned Stokes~$I$ continuum image (using all baselines with `Briggs 0.5' weighting scheme) corresponding to the LST:23.5h bin. We observe that subtraction of the bright sources Cas\,A and Cyg\,A leaves residuals at the $\sim1-2\%$ level. 

We also produced another set of higher resolution Stokes~$I$ snapshot images of the calibrated data with 1~min integration per snapshot, for every night used in the analysis, to study the ionospheric condition during these observations. A lower baseline cut is applied to avoid the large-scale Galactic diffuse emission. A source database was created by selecting $\sim 2500$ compact sources from the combined image of all nights using \textsc{pybdsf} software \citep{mohan2015}. The sources from the database were matched in snapshot images using \textsc{pybdsf} and position shifts corresponding to 650 bright sources (out of 2500) were obtained. These position shifts are used in later sections to assess ionospheric conditions for different nights.

\section{Noise statistics}\label{sec:Noise-stat}

We derive noise statistics of the data using time differenced Stokes~$V$ visibilities ($\delta_t \mathcal{V}_{V}(u,v,\nu)$). The calibrated visibilities are divided into even and odd samplings at 4\,s time resolution and gridded. At this time cadence, the sky (circularly polarised emission), ionospheric effects, and the PSF do not vary appreciably, cancelling out in the difference. Thus, the difference between even and odd gridded Stokes~$V$ visibilities provides a reasonable estimate of the thermal noise (apart from a $\sqrt{2}$ factor). This estimate may be used to obtain System Equivalent Flux Density (SEFD) by rearranging the following equation \citep{thompson2001}:
\begin{equation}\label{eqn:vis-noise}
\begin{split}
&  \sigma_{\text{vis}}(u,v,\nu) = \dfrac{\sqrt{2}k_{B}T_{\text{sys}} }{A_{\text{eff}}N_{\text{vis}}(u,v,\nu)\sqrt{\Delta\nu\Delta t}}, \ \text{and} \\
& \text{SEFD} = \dfrac{2k_{B}T_{\text{sys}}}{A_{\text{eff}}},
\end{split}
\end{equation}

where $\sigma_{\text{vis}}$ and $N_{\text{vis}}$ are visibility noise and number of visibilities per $uv$-cell, respectively. Also, $\Delta\nu$ and $\Delta t$ are frequency channel width and time resolution, respectively. We found the average SEFD values for LST bins~1 and~2 to be $\approx 1.93\,$MJy ($\pm 33\,$kJy) and $\approx 1.95\,$MJy ($\pm 49\,$kJy), respectively. These estimates are similar (within a few per cent) to other SEFD estimates for AARTFAAC imager ($\approx 2\,$MJy\footnote{AARTFAAC team via private communication}). Differencing Stokes~$V$ visibilities corresponding to adjoining channels ($\delta_{\nu} \mathcal{V}_{V}(u,v,\nu)$) also provides an estimate of frequency uncorrelated noise. We compare the ratio of frequency averaged angular power spectra ($\langle C_{\ell}[\delta_{\nu} \mathcal{V}_{V}] \rangle / \langle C_{\ell}[\delta_t \mathcal{V}_{V}]\rangle$) of time and frequency difference Stokes~$V$ visibilities. Figure \ref{fig:Noise-PS-ratio} shows the ratio for eight nights in the two LST bins. We observe that the ratio varies between $1.0-1.2$ for different nights in either LST bins. However, the mean of ratios for different nights varies between $1-1.1$ and shows a weak baseline dependence. We suspect that the excess is due to the residual part of the Stokes~$I$ sky leakage to Stokes~$V$ (at 0.2\% level at 74~MHz in Stokes~$I$) in channel difference visibilities which is coherent over different nights. The residual sky is small enough and is of the order of the thermal noise for a single time-slice, however, appears in the incoherent mean of the ratio. The baseline dependence might be caused by the scale dependence of the residual sky emission. 

We use channel difference visibilities as a proxy for frequency uncorrelated noise and time difference visibilities as a proxy of thermal noise in the data. Previous LOFAR-EoR data analyses in \cite{patil2017,gehlot2019} used Stokes~$V$ data itself as a noise estimator because only a tiny fraction of sky is circularly polarised making Stokes~$V$ a proxy of thermal noise of the system. However, in wide-field arrays such as AARTFAAC, the Stokes~$I$ to Stokes~$V$ polarization leakage can become more significant, contaminating the otherwise clean Stokes~$V$ data. Another thermal noise estimator is time-differenced Stokes~$V$ visibilities as described above; however, it does not account for uncorrelated errors in the frequency direction. Therefore, we simulate noise using equation~\ref{eqn:vis-noise}, with SEFD estimates based on channel differenced Stokes~$V$ visibilities. Associated noise visibilities $\mathcal{V}_{\text{N}}(u,v,\nu)$ are later used to correct for the noise bias in Stokes~$I$ residual power spectra. This approach is similar to the one used by \cite{mertens2020} to simulate noise visibilities. 
 
 \begin{table*}
    \centering
    \begin{tabular}{l|c|c|c|c}
        \hline
        \textbf{Covariance Model} & \textbf{Hyperparameters} & \textbf{Prior} & \textbf{MCMC estimate} & \textbf{MCMC estimate}\\
          & & & \textbf{(LST bin~1)} &  \textbf{(LST bin~2)} \\
        \hline
         Intrinsic foregrounds (${\bf K}_{\text{int}}$) & $\eta_{\text{int}}$ & $+\infty$ & - & - \\
         \vspace{1mm}
         & $\sigma_{\text{int}}^2/\sigma_{\text{n}}^2$ & - & $562.26_{-10.61}^{+6.59}$ & $521.06_{-8.18}^{+8.51}$\\
         \vspace{1mm}
         & $l_{\text{int}}$ & $\mathcal{U}(5,100)$  & $>77.05$ & $>67.81$\\
         \vspace{1mm}
         Instrumental Mode mixing (${\bf K}_{\text{mix}}$) & $\eta_{\text{mix}}$ & $3/2$ & - & - \\
         \vspace{1mm}
         & $\sigma_{\text{mix}}^2/\sigma_{\text{n}}^2$ & - & $119.54_{-1.12}^{+2.46}$ & $116.91_{-0.76}^{+2.72}$\\
         \vspace{1mm}
         & $l_{\text{mix}}$ & $\mathcal{U}(0.5,20)$ & $1.26_{-0.006}^{+0.009}$ & $1.26_{-0.004}^{+0.012}$\\
         \vspace{1mm}
         Sub-band Ripple (${\bf K}_{\text{SB}}$) & $\eta_{\text{RBF}}$ & $+\infty$ & - & - \\
         \vspace{1mm}
         & $\sigma_{\text{RBF}}^2/\sigma_{\text{n}}^2$ & - & ${9.61}_{-0.031}^{+0.093}$ & ${9.137}_{-0.052}^{+0.064}$ \\
         \vspace{1mm}
         & $l_{\text{cos}}$ & $\mathcal{U}(0.02,0.05)$ & $0.031_{-0.000004}^{+0.000003}$ & $0.031_{-0.000005}^{+0.000003}$\\
         \vspace{1mm}
         The 21-cm signal (${\bf K}_{\text{21}}$) & $\eta_{\text{21}}$ & $1/2$ & - & - \\
         \vspace{1mm}
         & $\sigma_{\text{21}}^2/\sigma_{\text{n}}^2$ & - & $<0.0054$ & $<0.0068$\\
         \vspace{1mm}
         & $l_{\text{21}}$ & $\Gamma(7.2,8.5)$ & $>0.43$ & $>0.52$ \\
    \end{tabular}
    \caption{List of hyperparameters, corresponding priors and MCMC estimates (for 8~nights combined data) for different covariance components in the final GP model.}
    \label{tab:GP-covariance}
\end{table*}
 
\subsection{Combining multiple time-slices}\label{subsec:combine_nights}

In an ideal case, the number of visibilities per gridded $uv$-cell ($N_{vis}(u,v,\nu)$) would represent the visibility weight if noise on each visibility follows the same noise statistics. \cite{mertens2020} used a methodology to account for the night to night variations in the data by calculating modified visibility weights as:

\begin{equation}\label{eqn:vis-weights}
    W_{V}(u,v) = \dfrac{1}{\text{MAD}(\delta_{\nu} \mathcal{V}_{V}(u,v,\nu))\sqrt{N_{\text{vis}}(u,v,\nu)}} .
\end{equation}

This equation computes weights using a robust Median Absolute Deviation (MAD) estimate from channel difference Stokes~$V$ visibilities. These weights reflect night-to-night variations and baseline dependence in the otherwise (theoretically) invariant per-visibility noise. Following the method in \cite{mertens2020}, to increase the robustness of the estimator, the baseline profile of weights $W_{V}(|\vect{u}|)$ is fitted with a 3rd-order polynomial to obtain $\widehat{W}_{V}(|\vect{u}|)$ that is further normalised such that $\langle \widehat{W}_{V}(|\vect{u}|)\rangle_{(|\vect{u}|,m)} \equiv 1$ after averaging over all baselines and time-slices ($m$) in a given LST bin. $\widehat{W}_{V}(|\vect{u}|)$ can be used to obtain visibility weights per night as

\begin{equation}\label{eqn:final-weights}
\mathcal{W}(u,v,\nu) = N_{\text{vis}}(u,v,\nu) \widehat{W}_{V}(|\vect{u}|)    
\end{equation}

and different time-slices within an LST bin are then combined using a weighted average as

\begin{equation}\label{eqn:combine-vis}
\mathcal{V}_{m} (u,v,\nu) = \dfrac{\sum_{i=1}^{m}\mathcal{W}_{i} (u,v,\nu) \mathcal{V}_{i} (u,v,\nu)}{\sum_{i=1}^{m}\mathcal{W}_{i} (u,v,\nu)}, \\
\end{equation}
 
 where $\mathcal{V}_{i} (u,v,\nu)$ are the visibilities corresponding to $i$-th time-slice (note that all time-slices in a given LST bin have the same phase centre) and $\mathcal{V}_{m} (u,v,\nu)$ are the visibilities corresponding to $m$ time-slices combined. In the previous section, we observed that these night to night variations in the frequency uncorrelated noise are relatively small. Therefore, Stokes~$I$ visibilities are combined using $N_{\text{vis}}(u,v,\nu)$ as weights. However, we use this weight definition to define weights for the noise visibilities $\mathcal{V}_{\text{N}}(u,v,\nu)$ and use inverse variance weighting to combine the noise visibility cubes optimally.

\section{Foreground removal}\label{sec:GPR_fg_removal}

Subtraction/isolation of the bright foreground emission is a crucial step in 21-cm signal experiments. The intrinsic foreground emission has two dominant components viz. diffuse emission (Galactic synchrotron and thermal emission), and extra-galactic sources (e.g. radio galaxies, clusters and supernova remnants) \citep{dimatteo2002,zaldarriaga2004,bernardi2009,ghosh2012}. In addition to this, the instrument imparts spectral structure on the data called instrumental mode-mixing due to its frequency response \citep{datta2010,morales2012,trott2012,vedantham2012,hazelton2013}. On the other hand, the 21-cm signal varies rapidly with frequency. Gaussian Process Regression (GPR) \citep{rasmussen2005} exploits this distinct spectral behaviour of the intrinsic foregrounds, instrumental mode-mixing, and the 21-cm signal to separate them from each other. GPR models these different components with Gaussian Processes (GPs), using different covariance functions representing the spectral correlation functions of the different components. Readers may refer to \cite{mertens2018,mertens2020} for an overview of GPR and its application for foreground removal and signal separation. 

In the current analysis, we use DD-calibration to remove only two bright sources Cas\,A and Cyg\,A unlike the DD-calibration in LOFAR beam-formed data analysis where several directions are used in DD-calibration to remove compact sources in the sky-model (see, e.g. \citealt{patil2017,gehlot2019,mertens2020}). The reason behind choosing this strategy is the fact that individual dipoles are less sensitive and have wider FoV than beam-formed stations rendering the DD-calibration (with several directions) on AARTFAAC data unfeasible from the standpoint of obtaining enough signal-to-noise ratio towards every direction and very high computational requirements due to a large number of antenna elements. Therefore, we tune GPR to remove diffuse+compact foreground emission and the instrumental mode mixing component. We select 42 channels of 65.1~kHz width (totalling 2.73~MHz bandwidth) to perform the foreground removal. 

 \begin{figure*}
    \centering
    \includegraphics[width=1\textwidth]{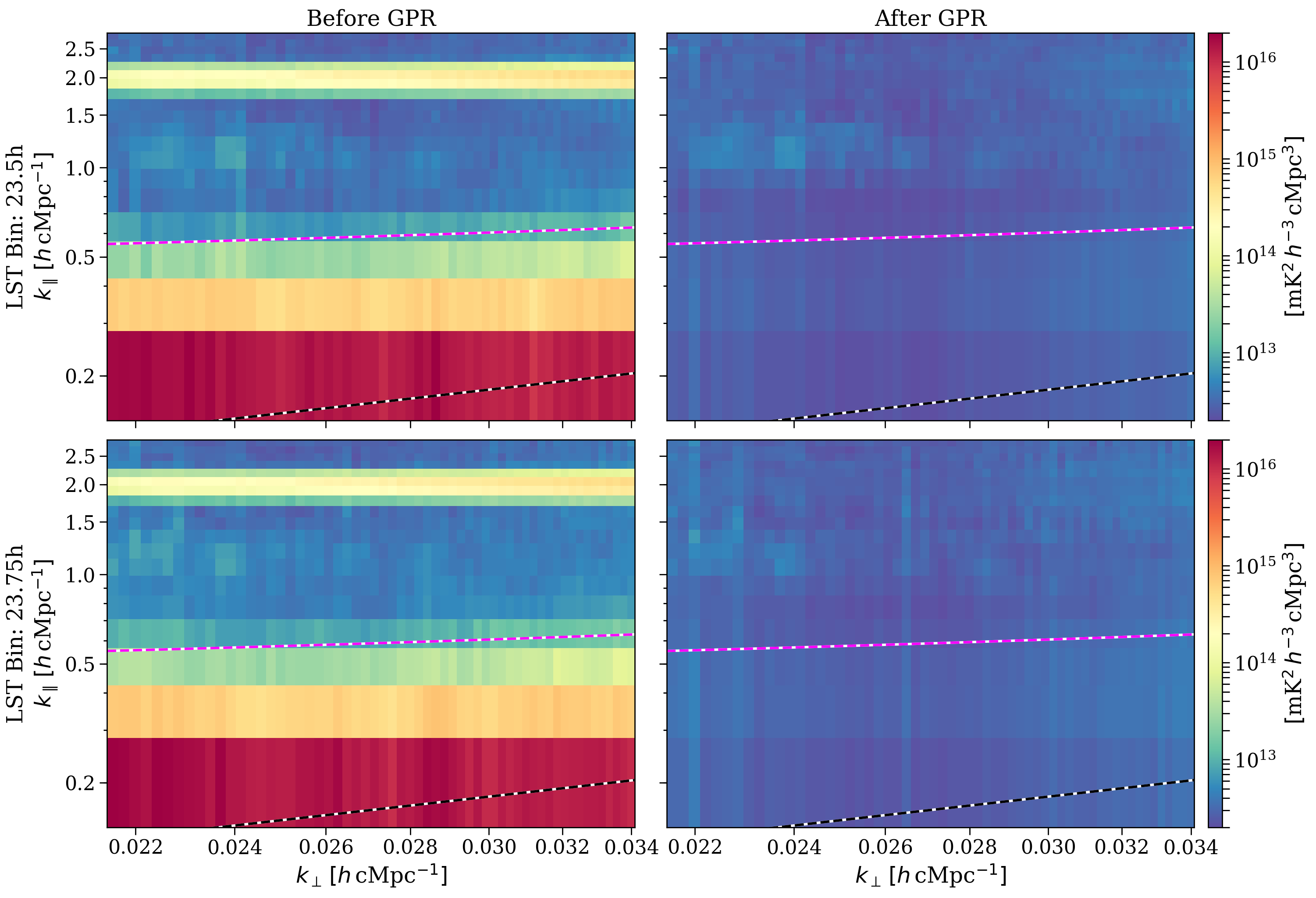}
    \caption{Cylindrically averaged Stokes~$I$ power spectra $P(k_{\perp},k_{\parallel})$ of combined power spectra of 8 nights per LST bin before (left column) and after (right column) foreground removal with GPR. Top and bottom rows correspond to LST:23.5h and LST:23.75h, respectively. Black dashed line corresponds to the instrumental horizon, and the purple dashed line corresponds to horizon buffer accounting for the window function.}  
    \label{fig:PS2D_I_combined}
    \end{figure*}

\subsection{Covariance model selection}\label{subsec:covariance_model}

The observed data $\vect{d}$ can be modelled as the sum of foreground components (intrinsic and mode-mixing) $\vect{f}_{\text{fg}}(\nu)$ that are coherent over the wide frequency range (intrinsic foregrounds coherent scale >10 MHz, mode mixing component coherent scale 1-5 MHz ), the 21-cm signal $\vect{f}_{\text{21}}(\nu)$ which is expected to decorrelate for > 1~MHz scales, and uncorrelated noise component ($\vect{n}$), i.e.

\begin{equation}\label{eqn:gpr-model}
    \vect{d} = \vect{f}_{\text{fg}}(\nu) + \vect{f}_{\text{21}}(\nu) + \vect{n}
\end{equation}

The covariance of the above GP model ($\textbf{K} = \textbf{K}_\text{fg} + \textbf{K}_\text{21} + \textbf{K}_\text{n}$) is composed of covariances of different components in the model, e.g. foregrounds, the 21-cm signal, and noise.  Selection of covariance functions for the final covariance model is driven by data in a Bayesian framework where the model that maximises the evidence is chosen. A covariance model that matches the data allows us to obtain an estimation of the power spectrum. We use Matern class covariance function ($\kappa_{\text{Matern}}$) \citep{stein1999} to represent the covariance of different components of the GP model.
\begin{equation}\label{eqn:matern}
\kappa_{\text{Matern}} (\nu_{\text{p}},\nu_{\text{q}}) = \sigma^2 \dfrac{2^{1-n}}{\Gamma (n)} \left( \dfrac{\sqrt{2n}|\nu|}{l} \right)^{n} K_{n} \left( \dfrac{\sqrt{2n}|\nu|}{l} \right) \ ,
\end{equation}
where $|\nu| = |\nu_{\text{q}} - \nu_{\text{p}}|$ is the absolute frequency separation between two channels, $\sigma^2$ is the variance, and $K_{n}$ is the modified Bessel function of the second kind (not to be confused with covariance matrices). The coherence scale is set using the `hyper-parameter' $l$. Listed below are the covariance models for various components.

\begin{enumerate}
    \item \textit{Intrinsic foregrounds}$-$ The intrinsic sky emission such galactic diffuse emission (synchrotron, free free emission etc.), extra-galactic sources (e.g. radio galaxies and clusters, supernova remnants) compose intrinsic foregrounds. These foregrounds tend to be smooth at frequency scales $\gtrsim 10$~MHz. For intrinsic foreground component (${\bf K}_{\text{int}}$), we set $n=+\infty$, which yields a Radial Basis Function (RBF) (equivalent to a Gaussian covariance function) and set uniform prior $\mathcal{U}(5,100)$~MHz for the frequency coherence scale $l_{\text{int}}$. 
    \\
    \item \textit{Instrumental mode-mixing}$-$ Chromatic behaviour of an instrument such as Instrumental bandpass (and poly-phase filter passband), cross-talk/mutual coupling between receivers impart less-smooth spectral structure onto otherwise smooth intrinsic foreground emission. Moreover, residuals due to imperfect calibration are also chromatic in nature. These effects combined together from the mode-mixing component. We use the Matern covariance function with $n=3/2$, with a uniform prior $\mathcal{U}(0.5-20)$~MHz to model the mode-mixing covariance (${\bf K}_{\text{mix}}$). Furthermore, frequency sub-bands in AARTFAAC also affected by a frequency structure due to the polyphase filter bank that repeats after every 3~channels (195.3~kHz, referred to as sub-band ripple, hereafter). We model the covariance of this sub-band ripple (${\bf K}_{\text{SB}}$) using a product of a radial basis function and a cosine covariance function. The latter is written as
    \begin{equation}\label{eqn:cosine}
        \kappa_{\text{Cosine}}(\nu_p,\nu_q) = \cos\left(|\nu|/l_{\cos}\right)
    \end{equation}
    where $l_{\cos}$ is the lengthscale of the cosine function with period $p = l_{\cos}/2\upi$. Coherence length-scale of RBF covariance function is fixed at 1.5~MHz and a uniform prior $\mathcal{U}(0.02,0.05)$~MHz used for the lengthscale of the cosine function.
    \\
    \item \textit{The 21-cm signal}$-$ So far any information about 21-cm signal fluctuations comes from simulations, due to the lack of a detection. The covariance shape of the signal is also unknown. However, simulations of the CD and EoR (e.g. \textsc{21cmfast} simulations; \citealt{mesinger2011}) may be used to understand covariance properties of the expected 21-cm signal, which is expected to decorrelate at $\gtrsim 1$~MHz scales. \cite{mertens2018} used \textsc{21cmfast} simulations to show that the exponential covariance function well approximates the frequency covariance of 21-cm signal from different phases of the Cosmic Dawn and Reionization Epoch. Therefore, we choose exponential covariance function to represent 21-cm signal covariance (${\bf K}_{21}$) and is obtained by setting $n=1/2$ in equation \ref{eqn:matern} with $\sigma_{21}^2$ and $l_{21}$ as hyper-parameters and a Gamma distribution prior $\Gamma(\alpha,\beta)$ with $(\alpha,\beta) = (7.2,8.5)$. 
    \\
    
    \item \textit{The Noise}$-$  We simulate noise covariance (${\bf K}_{\text{sn}}$) using the same approach for simulating noise visibilites (using equation~\ref{eqn:vis-noise} with weights described by equation~\ref{eqn:vis-weights}).
    
\end{enumerate}

The final covariance model (${\bf K}$) is a sum of all the components described above and is given by
\begin{equation}
    {\bf K} = {\bf K}_{\text{int}} + {\bf K}_{\text{mix}} + {\bf K}_{\text{SB}} + {\bf K}_{21} + {\bf K}_{\text{sn}}.
\end{equation}
The final GP model consists of 8~hyperparameters that are optimised in a Bayesian manner by maximising the evidence using an MCMC approach. In addition to using a covariance kernel, we use Principal Component Analysis (PCA) in conjunction with GPR to remove the sub-band ripple from the data. First, GPR is used to remove the foregrounds. Next, PCA is run on the residuals, and the first principal component is subtracted from the original visibilities before the foregrounds were removed. Finally, GPR is performed on the residuals after subtraction of the first principal component to remove the remaining part of the sub-band ripple component. Since the sub-band ripple is independent of direction, the PCA technique is sufficient to mitigate the residual ripple post GPR. Moreover, it does not impact the 21-cm signal since the PCA component is the same for all baselines, and hence only removes instrumental effects such as bandpass errors. Foreground removal is performed separately on each time-slice in both LST bins, as well as different combinations of averaged time-slices. Table~\ref{tab:GP-covariance} lists the hyperparameters of the GP model, their priors used for GPR foreground removal and values (with marginalised errors) that maximise the evidence obtained using MCMC parameter estimation. We note that frequency coherence lengthscale ($l_{\text{int}}$) of intrinsic foregrounds is poorly constrained and hits the prior boundary (only lower limit is reported in table~\ref{tab:GP-covariance}) due to limited bandwidth. The 21-cm signal lengthscale also hits the prior boundary (which is expected); hence only the lower limit is reported. 

Similar to \cite{mertens2020}, a bias correction to the power-spectrum estimation is applied during the GPR foreground removal to obtain an unbiased estimate of covariance of residuals. This bias depends on the Dynamic Range (DR) of the data. We find that the normalised intrinsic foreground variance $\sigma_{\text{int}}^2/\sigma_{\text{n}}^2$ (a proxy of the DR of the data) for the two LST bins estimated during GPR is similar to the value reported in \cite{mertens2020} that corresponds to the LOFAR-HBA data after subtraction of compact sources. This is expected as the AARTFAAC data is significantly noisier than LOFAR-HBA data and has similar DR as that of LOFAR-HBA data post subtraction of compact sources. In addition, the inferred coherence length-scales obtained for the intrinsic foregrounds and mode-mixing component are $>1$~MHz. In contrast, the 21-cm signal is expected to decorrelate on coherence scales less than 1~MHz, showing that GPR optimally separates the foregrounds from the data without affecting the faint 21-cm signal.

\subsection{Impact of the ionosphere on foreground removal}\label{subsec:ionosphere}
Turbulence in ionospheric plasma introduces phase shifts in the electromagnetic wave-front propagating through the ionosphere. The phase shifts are dispersive and have a significant impact on the data observed at low frequencies. Full-sky images produced with AARTFAAC data allows accessing ionospheric information of the observations \citep{koopmans2010,vedantham2015,vedantham2016}. Although the DI calibration mitigates the average ionospheric distortion along the effective Line of Sight (LOS), the ionosphere and therefore its distortion varies over the field of view. A linear gradient in the electron density over the array, for a given direction on the sky, results in an apparent shift of the position of a source in the image towards that direction. To first order, the ionospheric structure is expected to be linear over the $\sim1.5$~km patch size formed by the array. However, this gradient varies depending on the LOS direction \citep{mevius2016}. We used the position shifts obtained from 1~min snapshot images as described in Section~\ref{subsec:Imaging} to investigate ionospheric variability during different nights used in the analysis. Projecting these position shifts on a virtual ionospheric layer provides a direct probe of the ionospheric disturbances (see, e.g. \citealt{loi2015,jordan2017}). The mean and variance of these source position shifts provide an initial estimate of the ionospheric conditions during the observations. We find the average position shifts to vary between $1.7-2.2$~arcmin, suggesting relatively mild ionospheric conditions. However, two nights show relatively higher variations, but we do not observe that results from these nights do not show any deviations from the results corresponding to the nights with relatively mild ionospheric conditions.

Moreover, we only use a small baseline range of $25-40\lambda$ for the foreground removal, that corresponds to an image resolution of the order of a degree. The average positional shifts caused by the ionosphere are a fraction of the resolution; hence, it does not significantly impact the GPR foreground removal. Additionally, the overall coherent integration time for power spectrum estimation is about 2~hours, and the residuals are mainly dominated by the thermal noise. However, ionospheric effects may start to play a role for deeper integrations, therefore studying these effects using the available data will itself be a subject of further study, once more data is analysed.

\subsection{Power Spectrum estimation}\label{subsec:power-spectra}
After having removed all FGs, we estimate the PS from the residual data cubes. For a given survey of co-moving volume $\mathbb{V}$, the power spectrum $P(\vect{k})$ of a brightness temperature field $T(\vect{r})$ is defined as 
\begin{equation}
    P(\vect{k}) = \mathbb{V} \langle |\tilde{T}(\vect{k})|^2 \rangle,
\end{equation}
where $\tilde{T}(\vect{k})$ is the discrete Fourier transform of $T(\vect{r})$:
\begin{equation}
    T(\vect{k}) = \dfrac{1}{N_x N_y N_{\nu}} \sum_{\vect{r}} T(\vect{r}) \, \text{e}^{2\upi i \vect{k}\cdot\vect{r}} \,.
\end{equation}
The wave-vector $\vect{k}$ has components $(k_x,k_y,k_{\parallel})$ and are defined as \citep{morales2004}:
\begin{equation}
    k_x = \dfrac{2\upi u}{D(z)}, \ k_y = \dfrac{2\upi v}{D(z)}, \ k_{\parallel} = \dfrac{2\upi\nu_{21}H_{0}E(z)}{c(1+z)^2}\eta 
\end{equation}
where $z$ is the redshift of observation, $\nu_{21}$ is the rest-frame frequency of the 21-cm transition, $D(z)$ is the transverse comoving distance, $H_0$ is the Hubble constant and  $E(z) \equiv \sqrt{ \Omega_m(1+z)^3 + \Omega_{\Lambda}}$ \citep{hogg1999}, and $\eta$ is the Fourier dual to frequency. The cylindrically averaged power spectrum $P(k_{\perp},k_{\parallel})$ may be obtained from $P(\vect{k})$ as:
\begin{equation}
P(k_{\perp},k_{\parallel}) = \langle P(\vect{k}) \rangle_{(k_x,k_y)}, \ \text{and} \ k_{\perp} = \sqrt{k_x^2 + k_y^2}\,.      
\end{equation}
Similarly, the spherically averaged dimensionless power spectrum ($\Delta^2(k)$) can be obtained from $P(\vect{k})$ as:
\begin{equation}
\Delta^2(k) = \dfrac{k^3}{2\upi^2} \langle P(\vect{k}) \rangle_{(k_x,k_y,k_{\parallel})}.
\end{equation}

We use the gridded visibilities (in temperature units) before and after foreground removal produce various power spectrum products.

\begin{figure*}
\centering
\includegraphics[width=1\textwidth]{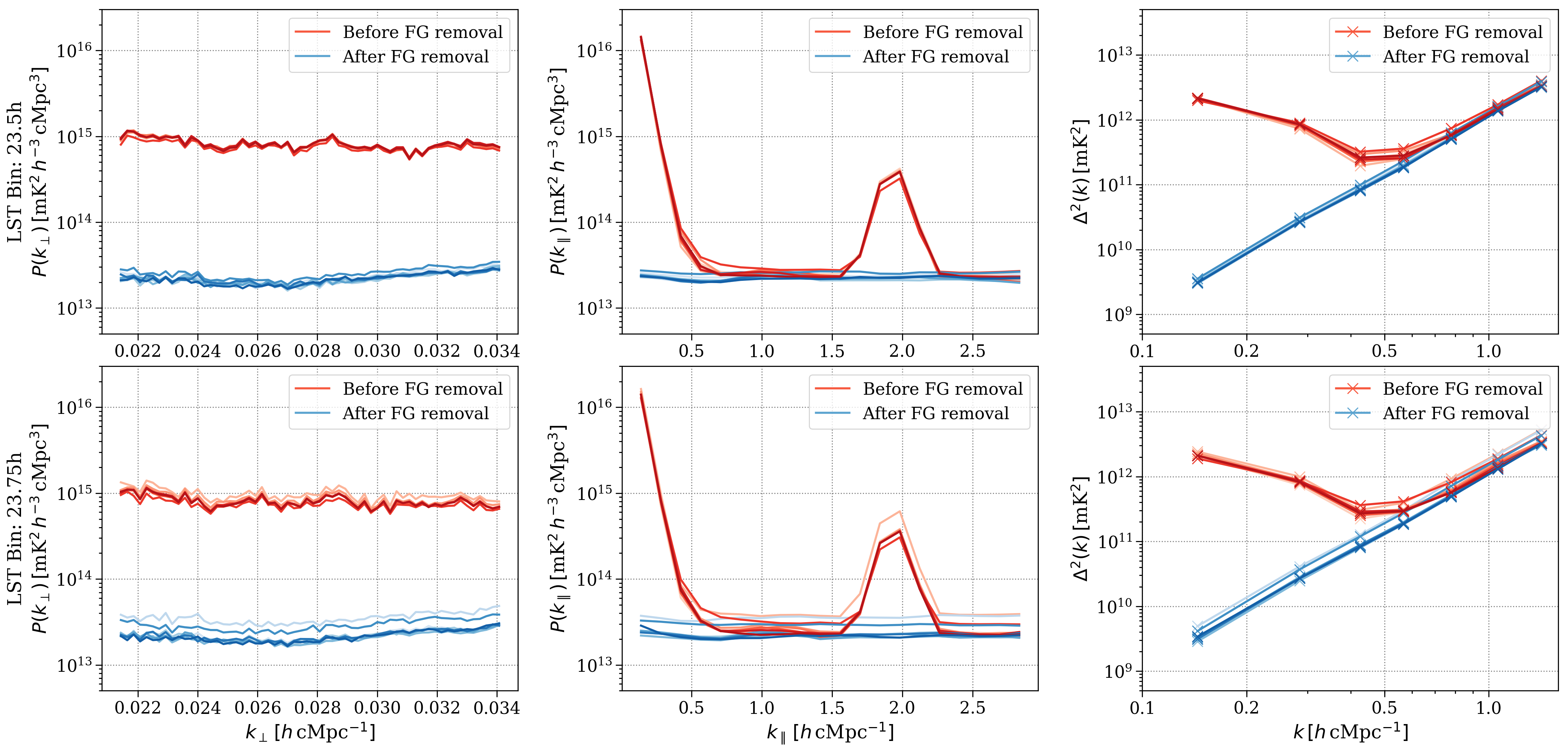}
    \caption{Stokes~$I$ power spectra before (reds) and after (blues) foreground removal of different time-slices used in the analyses. Left column shows $P(k_{\perp})$ which is the average of $P(k_{\perp},k_{\parallel})$ along the $k_{\parallel}$ axis. Middle column shows $P(k_{\parallel})$ which is the average of $P(k_{\perp},k_{\parallel})$ along the $k_{\perp}$ axis. Right column shows The spherically averaged power spectra. Different colour shades represent different time-slices.}  
\label{fig:PSmean}
\end{figure*}

\begin{figure*}
\centering
\includegraphics[width=\textwidth]{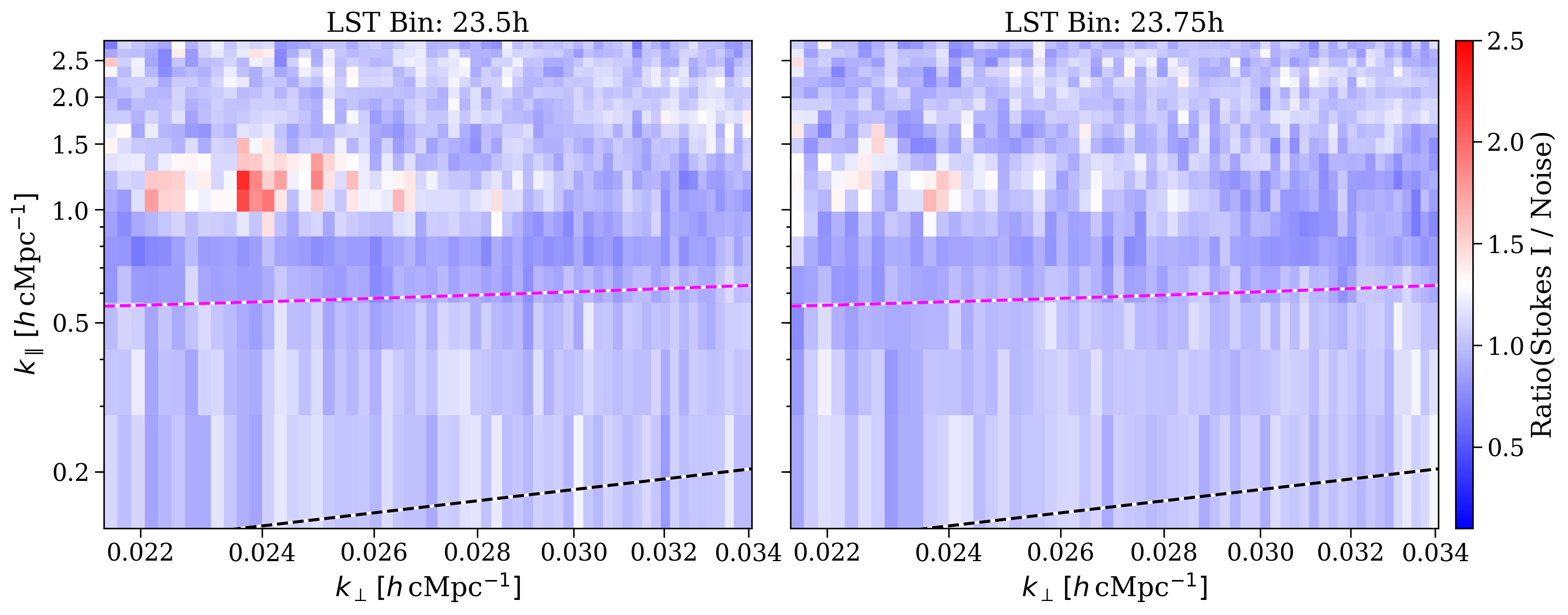}
    \caption{The ratio of residual Stokes~$I$ power spectrum after foreground removal and the estimated noise power spectrum. Left and right panels correspond to LST:23.5~h and LST:23.75~h, respectively. Black dashed line corresponds to the instrumental horizon, and the purple dashed line corresponds to horizon buffer accounting for the window function.} 
\label{fig:ps2d_ratio}
\end{figure*}

\begin{figure*}
\centering
\includegraphics[width=\textwidth]{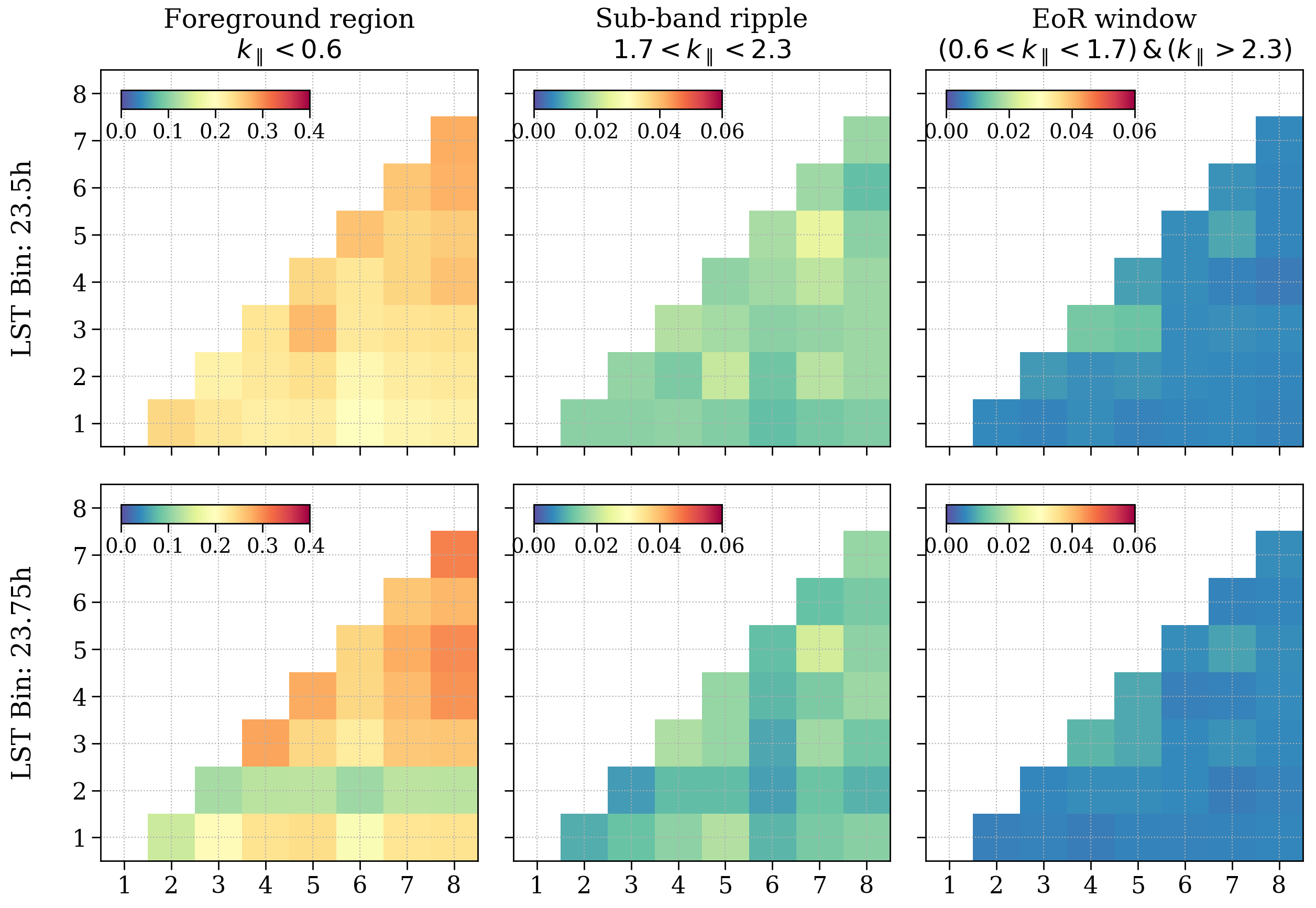}
    \caption{Mean cross-coherence of different pairs of time-slices in various regions (different columns) of $(k_{\perp},k_{\parallel})$ space. The three columns (left to right) correspond to the foreground, sub-band ripple and EoR window regions, respectively. The eight different time-slices in a given LST bin are numbered from 1~to~8. Top and bottom rows correspond to LST:23.5h and LST:23.75h bins, respectively.} 
\label{fig:coherence_combined}
\end{figure*} 

\section{Multi-night results}\label{sec:multi_night_results}

Here we discuss the results after processing and foreground removal for various time-slices in each LST bin.

\subsection{Power spectra results}\label{subsec:PS_results}

Cylindrically averaged power spectrum in $(k_{\perp},k_{\parallel})$ space is the most commonly used statistical tool to study the challenges associated with foreground contamination and systematic biases \citep{bowman2009,vedantham2012}. The wave mode $k_{\perp}$ represents the scale of the brightness temperature fluctuations in the plane perpendicular to the line of sight and the wave mode $k_{\parallel}$ represents the scale of the fluctuations along the line of sight. Foregrounds, the ionospheric effects and systematic biases which are smooth in frequency reside within a region often called ``the wedge''. 

Figure \ref{fig:PS2D_I_combined} shows the cylindrically averaged Stokes~$I$ power spectra $P(k_{\perp},k_{\parallel})$ of 8 time-slices combined datacubes (for the two LST bins) before and after foreground removal. The structure around $k_{\parallel}\sim 2.0~h\,\text{cMpc}^{-1}$ in power spectra before foreground removal is a leftover of the sub-band ripple. We observe that the foreground emission that originally dominated the lowest $k_{\parallel}$ modes in Stokes~$I$ power spectra, as well as the 195~kHz ripple, are effectively removed by GPR foreground removal. We still observe a faint structure between $1.0 \lesssim k_{\parallel} \lesssim 1.5~h\,\text{cMpc}^{-1}$ and $k_{\perp}\lesssim 0.026~h\,\text{cMpc}^{-1}$. Figure~\ref{fig:PSmean} shows averaged $P(k_{\perp},k_{\parallel})$ (along $k_{\parallel}$ and $k_{\parallel}$ axes, respectively) and spherically averaged power spectra $\Delta^2(k)$ for individual time-slices in the two LST bins. Power spectra for different time-slices in a given LST bin are similar on approximately all $k$ modes present in the data. Power spectra before and after foreground removal also have similar noise floors. The LST:23.75h bin has 1-2 nights with slightly higher noise floor, which is probably due to relatively worse data quality or calibration compared to the rest. The faint structure observed in combined power spectra in figure~\ref{fig:PS2D_I_combined} is at or below the noise level of individual time-slices (or possibly absent in some). However, it shows up in final combined data. Furthermore, it appears to be transient and does not correlate from night to night (discussed later). The structure is possibly faint RFI which remains undetected by the RFI mitigation strategy we have utilised.  
 
Figure~\ref{fig:ps2d_ratio} shows the ratio between the residual Stokes~$I$ power spectrum after foreground removal (shown in the right column of figure~\ref{fig:PS2D_I_combined}) and the corresponding noise power spectrum for the two LST bins. We observe that the ratio is approximately flat except for the faint RFI structure, as mentioned previously. For the LST:23.5h bin, the ratio has a Median of~$\sim1.01$ and a Median Absolute Deviation (MAD) of~$\sim0.08$. For the LST:23.75h bin, the Median$\sim1.00$ and MAD$\sim0.08$. This shows that Stokes~$I$ residuals are almost entirely noise dominated for both LST bins.

\begin{figure}
\centering
\includegraphics[width=\columnwidth]{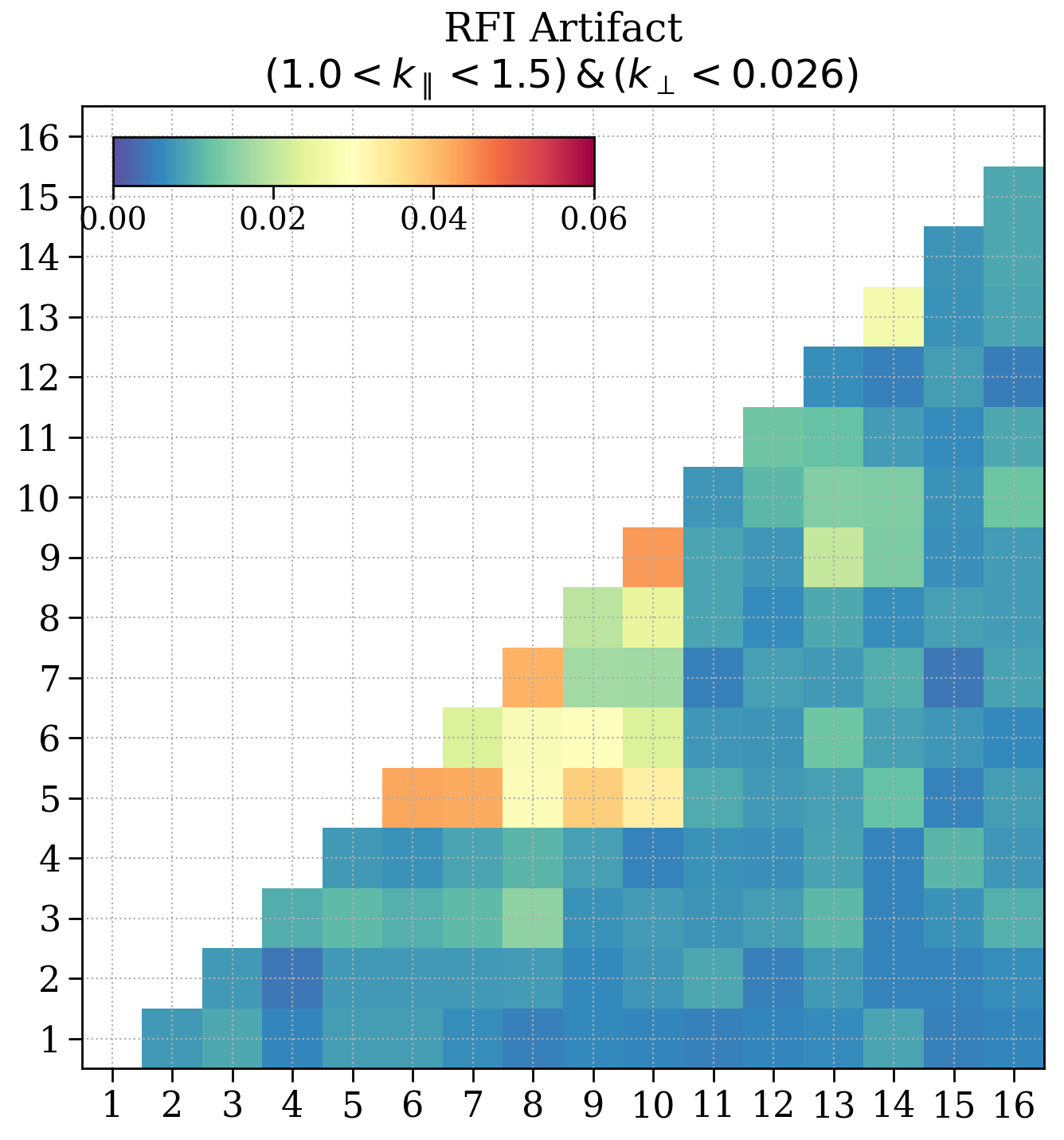}
    \caption{Mean cross-coherence of different pairs of time-slices in the region in $(k_{\perp},k_{\parallel})$ space surrounding the RFI structure. Odd-numbered time-slices correspond to LST:23.5h bin and even-numbered time-slices correspond to LST:23.75h bin (respectively). Time-slice pairs ($2i-1,2i$) where $i\in[1,8]$ belong to the same night except pair (3,4). Time-slices are arranged in the order of increasing observing date.} 
\label{fig:coherence_artifact}
\end{figure} 

\subsection{Cross-coherence between nights}\label{subsec:cross_coherence} 

Cross-coherence (or normalised cross-spectra) is a useful tool to understand correlations between different datasets, i.e. time-slices/LSTs in our case. We use cross-coherence further to study the correlation between residual data for different time-slices to better understand about residual foreground emission and other structures in cylindrical power-spectra. We use the definition in \cite{mertens2020} to define cross-coherence between two datasets in $(k_{\perp},k_{\parallel})$ space:

\begin{equation}
    C_{i,j}(k_{\perp},k_{\parallel}) = \dfrac{\langle |\tilde{T}_i^*(\vect{k})\tilde{T}_j^*(\vect{k})|^2 \rangle}{\langle |\tilde{T}_i(\vect{k})|^2 \rangle \langle |\tilde{T}_j(\vect{k})|^2 \rangle}\, ,
\end{equation}
where $\tilde{T}_i(\vect{k})$ is the Stokes~$I$ temperature cube in $\vect{k}$ space.  The value of $C(k_{\perp},k_{\parallel})$ can vary between $0-1$, representing no-correlation or maximum correlation, respectively.

We use residual Stokes~$I$ data to compute cross-coherence between different pairs of time-slices corresponding to different nights within an LST bin. We calculate the mean cross-coherence in 3~regions (as in \citealt{mertens2020}):
\begin{enumerate}
    \item The foreground region ($k_{\parallel}<0.6$), 
    \item The Sub-band ripple region ($1.7<k_{\parallel}<2.36$),
    \item The `EoR window' ($0.6<k_{\parallel}<1.7 \& k_{\parallel}>2.3$).
\end{enumerate}
Figure~\ref{fig:coherence_combined} shows the mean cross-coherence between different pairs of nights for a given LST bin for these three regions of $(k_{\perp},k_{\parallel})$ space. Foreground region shows a mean coherence of $\sim0.2-0.3$ for the two LST bins. It is possible that the observed correlation originates due to the sky-emission outside the FoV ($\sim120\deg$) used in the analysis, and remained unsubtracted even after GPR foreground removal. The sub-band ripple region, on the other hand, shows a low coherence $C_{i,j}\sim 0.01-0.02$, suggesting that the ripple is mitigated reasonably well with the strategy we have employed. However, it may become more severe as we integrate more data in future analyses. The coherence $C_{i,j}< 0.01$ of most pairs in the EoR window; however, time-slice pairs (3,4), (3,5) and (4,5) show slightly higher coherence ($C_{i,j}\sim0.01$). These time-slices are affected more by the faint RFI structure that also affects the coherence in the EoR window.

To understand the RFI structure better, we also compute the mean cross-coherence in the region $1.0<k_{\parallel}<{1.5} \ \& \ k_{\perp}<0.026$, which was affected the most by the RFI structure. For this case, we compute the coherence between all time-slices across the two LST bins. Figure~\ref{fig:coherence_artifact} shows the corner plot of the coherence of the RFI affected region of $(k_{\perp},k_{\parallel})$ space. We find that combinations of time-slice pairs between (5-10) show relatively higher coherence ($C_{i,j}\sim0.02-0.04$) and these nights were observed over 12 days. Note that time-slice pairs ($2i-1,2i$) belong to same $i$th observing night (except (3,4)). Any sky emission would decorrelate over short periods (e.g. between the two LST bins), whereas an RFI source on the ground would still be correlated over longer times. This provides additional evidence that the structure is transient and persisted over a span of 2-3 weeks, possibly some temporary source of local RFI. Other pairs show smaller correlations, similar to correlation levels in the EoR window.  

\begin{figure*}
\centering
\includegraphics[width=\textwidth]{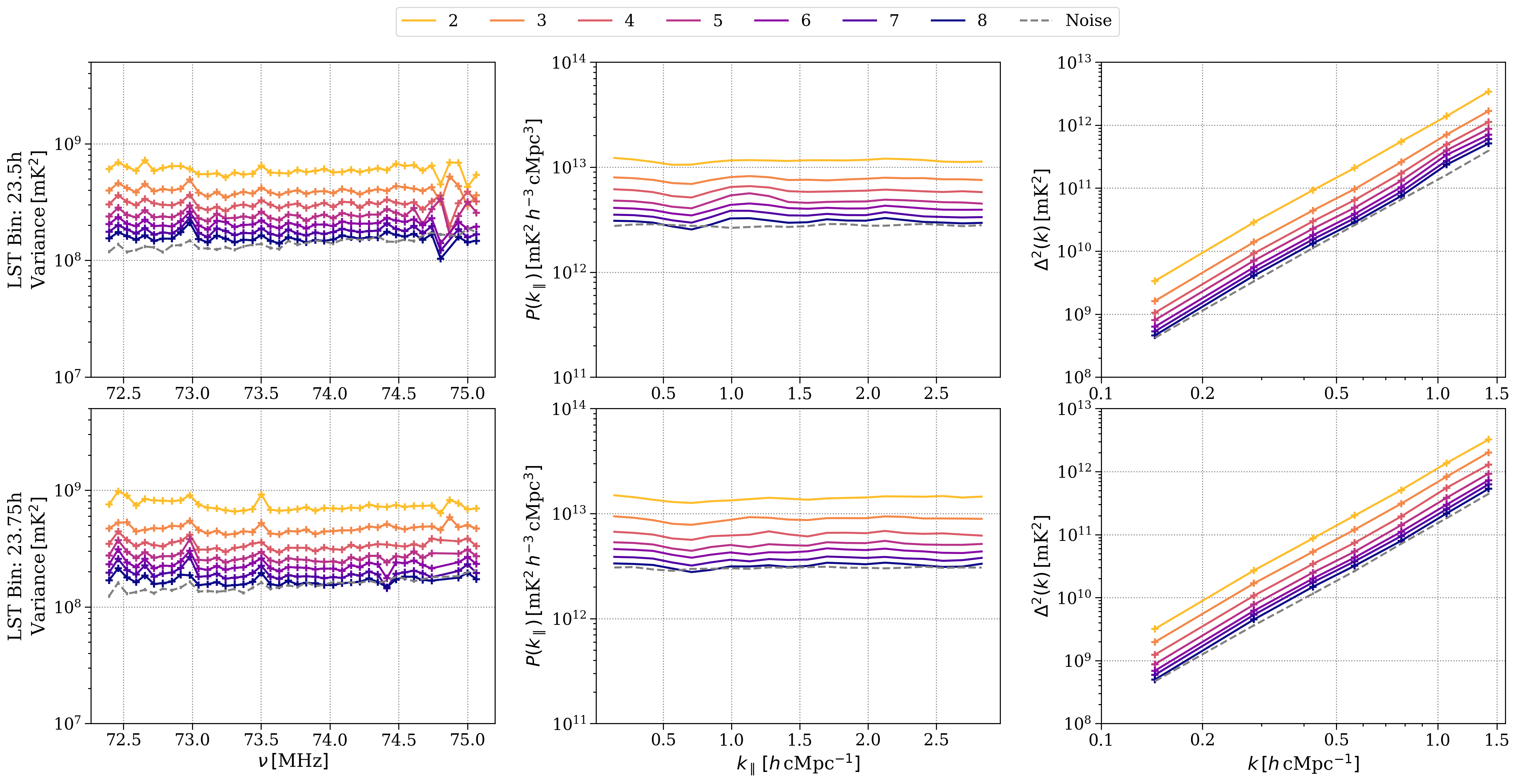}
    \caption{Various statistics for residual Stokes~$I$ of intermediate datasets with the increasing number of time-slice integration. Left to right: Variance, cylindrical power spectrum averaged over all $k_{\perp}$ modes, and spherically averaged power spectrum. Top and bottom rows correspond to LST:23.5h and LST:23.75h bins. Different colours correspond to the number of nights averaged (in increasing order from yellow to purple) in order of observing dates. The dashed grey line shows the thermal noise corresponding to 8 time-slices combined in a single LST bin. Note that spherically averaged power spherical power spectra shown here are not corrected for the noise bias.} 
\label{fig:multi_ps_combined}
\end{figure*} 

\begin{figure}
\centering
\includegraphics[width=\columnwidth]{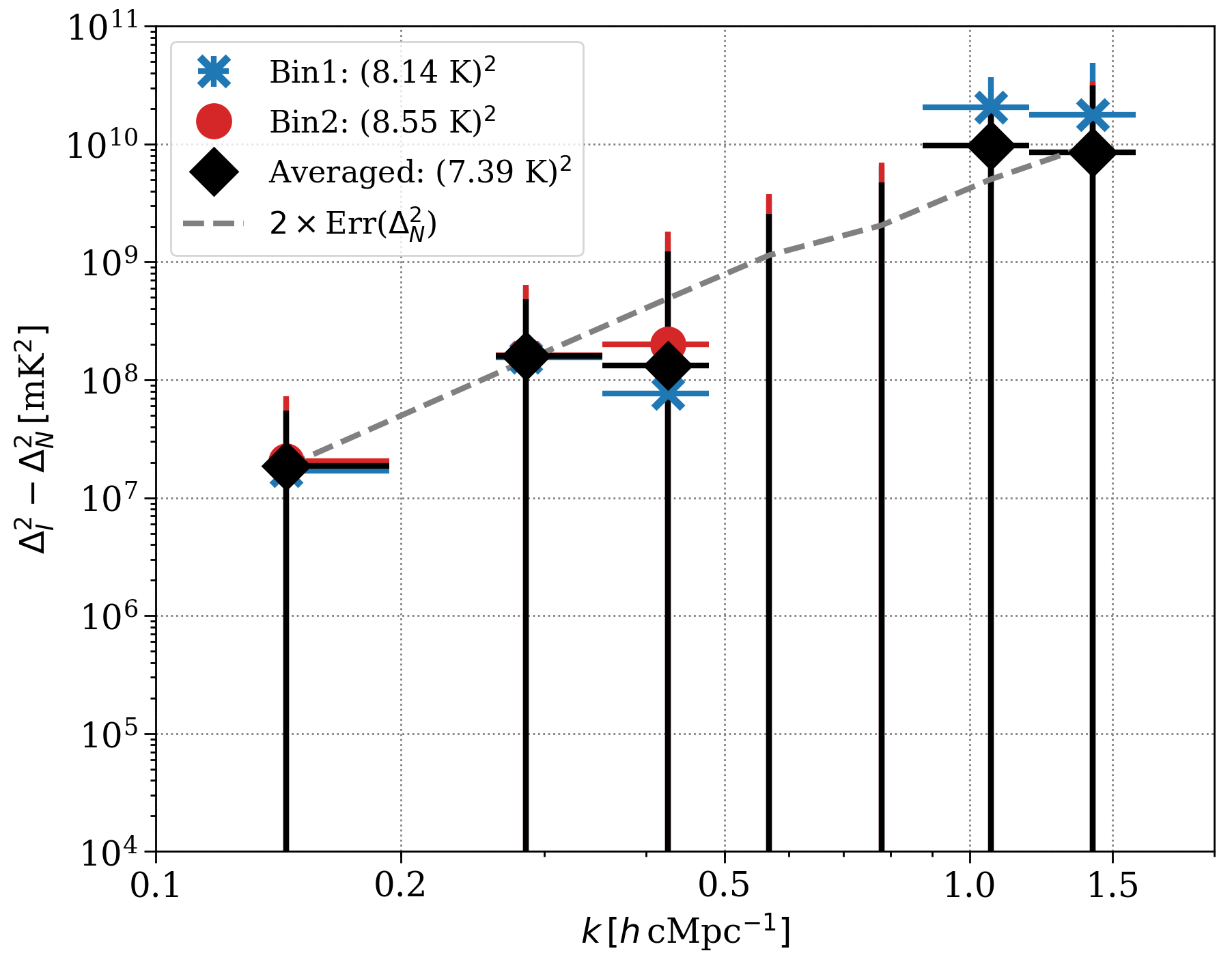}
    \caption{Noise bias corrected power spectra $\Delta_{21}^2$ for the two LST bins (`crosses' correspond LST:23.5h bin and `circles' correspond LST:23.5h bin). The incoherently averaged power spectrum is shown using Diamond markers. The dashed line shows the error on the noise power spectrum, which corresponds to the theoretically achievable $2\sigma$ upper-limit in 2~h of coherent averaging. The x-errorbars represent the range of $k$ bins and y-errorbars represent $2\sigma$ errors on the power spectra.} 
\label{fig:ps3d_ulim}
\end{figure} 

\subsection{Power spectra of combined time-slices}\label{subsec:combined_ps_results} 

We used the procedure described in Section~\ref{subsec:combine_nights} to combine different nights/time-slices in a given LST bin. The GPR foreground removal is performed on intermediate datasets obtained after adding time-slice on by one. Figure~\ref{fig:multi_ps_combined} shows the spectral variance, cylindrical power spectra averaged over all $k_{\perp}$, and spherically averaged power spectra of residual Stokes~$I$ data in intermediate combined datasets. As expected, variance and power spectra scale down in amplitude with integration time as we add more data to a given LST bin. Additionally, power spectra appear to be dominated by noise and devoid of any obvious, coherent structures that may emerge after integrating more time-slices. There are several frequency channels with slightly higher covariance, which is probably due to slightly higher levels of RFI flagging than the rest. Two channels on the higher frequency end show relatively lower variance, which cannot be explained by higher RFI fraction. We are investigating the cause of the low-variance of these two channels, and how it impacts the analysis and results. 

\subsection{Noise bias corrected power spectrum}\label{subsec:21cm_results} 

We compute the spherically averaged dimensionless Stokes~ $I$ power spectrum ($\Delta^2(k)$) of the residuals after GPR foreground removal for the combined (8~time-slices) data in the two LST bins. We use the corresponding simulated noise cubes $\mathcal{V}_{\text{N}}(u,v,\nu)$ to obtain the noise power spectrum that is used to correct for the noise bias. The noise bias corrected power spectrum (and associated uncertainty) can be written as:

\begin{equation}
\begin{split}
&  \Delta_{21}^2 = \Delta_{I}^2 - \Delta_{N}^2, \ \text{and}\\
&  \Delta_{21,\text{err}}^2 = \sqrt{\Delta_{I,\text{err}}^4 + \Delta_{N,\text{err}}^4}\,.
\end{split}
\end{equation}

The uncertainty value $\Delta_{21,\text{err}}^2$ on the power spectrum $\Delta_{21}^2$ includes the sample variance (due to the number of individual uv-cells and effective field of view) and a contribution from the uncertainty on the hyperparameters of the GP model used in the analysis. However, the inferred values of most hyperparameters of the GP model are well constrained with very small uncertainty levels ($\lesssim 1\%$ as shown in table \ref{tab:GP-covariance}), except for the 21-cm signal, suggesting that the GPR error contribution to the final uncertainty on the power spectrum can be ignored. Similar findings have been reported in the application of GPR on LOFAR-HBA data \citep{mertens2020} as well as on HERA data \citep{ghosh2020}.
 
We also combine $\Delta_{21}^2$ for the two LST bins incoherently using inverse variance weighting to obtain incoherently averaged power spectrum. These noise bias corrected power spectra for the two LST bins, and the incoherently averaged power spectrum are shown in figure \ref{fig:ps3d_ulim}. After 2~h ($8\times15~$min) of integration per LST bin, we obtain a $2\sigma$ upper limit of $\Delta_{21}^2 < (8139~\text{mK})^2$ at $k = 0.144~h\,\text{cMpc}^{-1}$ for LST:23.5h bin, and $\Delta_{21}^2 < (8549~\text{mK})^2$ at $k = 0.144~h\,\text{cMpc}^{-1}$ for LST:23.75h bin, respectively, in the redshift range $z=17.9-18.6$. The incoherently averaged power spectrum yields $\Delta_{21}^2 < (7388~\text{mK})^2$ at $k = 0.144~h\,\text{cMpc}^{-1}$. We observe that the upper limit scales down by a factor of~1.1, compared to the expected factor of $\sqrt{2}$. We also observed that the power spectra are dominated by noise at all $k$-scales probed by ACE.

\section{Summary and Future work}\label{sec:summary}

In this work, we described the AARTFAAC Cosmic Explorer (ACE) program motivated by the reported detection of the deep absorption feature in sky averaged spectrum of the 21-cm signal during Cosmic Dawn by the EDGES collaboration \citep{bowman2018}. Main results of the paper are summarised below:

\begin{enumerate}
\item We demonstrate the successful end-to-end application of the ACE data-processing pipeline (which is adapted from LOFAR-EoR data processing pipeline) to ACE data, starting from preprocessing and calibration to power spectrum estimation after foreground removal.
\newline
\item We observe that the ratio of noise estimates from the channel and time-differenced Stokes~ $V$ visibilities varies between $1.0-1.2$ for most time-slices in the two LST bins. The mean ratio per LST bin shows weak baseline dependence which is possibly caused by residual Stokes~$I$ to $V$ leakage. We use channel differenced Stokes~$V$ noise as an estimator of frequency uncorrelated noise and later for noise bias correction.
\newline
\item Residual power spectra reach the expected noise level to within 1~per cent. Cylindrically averaged Stokes~$I$ power spectra exhibit a faint structure between $0.1<k_{\parallel}<1.5$ and $k_{\perp}<0.026$. This structure appears to be transient (as shown by the cross-coherence test) and affects certain baselines. We suspect this structure is possibly caused by faint RFI which remained undetected by flagging strategy we have employed in the analysis. Combining multiple nights decreases the power as expected, and corresponding power spectra of combined data do not show any obvious coherent emission.
\newline
\item Even though the noise bias corrected power spectrum is still dominated by noise, it is not regarded as a detection because the power levels within $2\sigma$ are 2-3 orders of magnitude higher than the expected signal and the power still decreases by adding more data. We thus obtain a $2\sigma$ upper limit of $\Delta_{21}^2 < (8139~\text{mK})^2$ (or equivalently $(8.14~\text{K})^2$) and $\Delta_{21}^2 < (8549~\text{mK})^2$ (or equivalently $(8.55~\text{K})^2$) at $k = 0.144~h\,\text{cMpc}^{-1}$ for LST:23.5h and LST:23.75h bins, respectively. The incoherently averaged power spectrum yields $\Delta_{21}^2 < (7388~\text{mK})^2$ (or equivalently $(7.39~\text{K})^2$) at the same $k$ value. These limits correspond to the redshift range $z=17.9-18.6$.

\end{enumerate}

Although, the upper limits are still at least two orders of magnitude higher than signals predicted by simulations, adding more data in future will allow us to improve these limits and possibly exclude or constrain various astrophysical models that may explain the 21-cm signal from the Cosmic Dawn.

\subsection{Future outlook and forecast}\label{subsec:future_work}

In this work, we demonstrated the successful application of the new ACE data-processing pipeline and effectively reaching the expected noise levels. However, most of the steps used in the analysis are still fairly rudimentary and require improvements. In the future, we plan to improve the processing and analysis by improving several aspects of the ACE data-processing pipeline such as:

\begin{enumerate}
\item Improving Direction Independent calibration of the data by including a detailed sky-model (compact sources and large-scale diffuse emission). It will also allow a better direction subtraction/peeling of bright sources, Cas\,A and Cyg\,A, to mitigate residuals post subtraction.
\newline
\item The sub-band ripple is a dominant systematic in ACE data. Currently, we use a covariance model in GPR (along with PCA) to mitigate it, which is sub-optimal. The sub-band ripple is a multiplicative effect as it is caused by the bandpass shape of the polyphase filterbank. We are exploring methods and strategies to mitigate the sub-band ripple during the calibration step rather than during the post-imaging steps.
\newline
\item The RFI removal strategy used by \textsc{aoflagger} in the current analysis is based on LOFAR-LBA RFI mitigation strategy, which may be sub-optimal for noisier ACE data. We plan to improve the current RFI-mitigation strategy to  work on ACE data optimally. We also plan to use a near-field imaging technique to pinpoint whether the cause of the faint RFI is due to sources on the ground. In addition to this, we plan to alternatively explore RFI mitigation techniques that use polarization information and directional statistics of RFI for RFI detection and removal \citep{yatawatta2020}.
\newline
\item Our current analysis is based on coherently averaging data in a given LST range. In the future, we plan to widen the LST range using map-making methods based on spherical harmonics techniques to achieve a lower noise floor.
\newline
\item Low-frequency wide-field observations are more prone to effects from polarization leakage and the ionosphere. We plan to study the effect of polarization leakage and the ionosphere in ACE observations and mitigate them if required.

\end{enumerate}

\noindent Incorporating the improvements as mentioned above to the data processing pipeline allows us to decrease the power spectrum levels further. Assuming that foregrounds and systematics are optimally mitigated/removed, a power spectrum sensitivity of $\Delta_{21}^2 < (1\,\textrm{K})^2$ may be reached in 150~h of integration and even lower levels with the 500~h of data in hand. This sub-Kelvin sensitivity is already at the level of some predictions of the 21-cm fluctuations during the Cosmic Dawn \citep{fialkov2018,fialkov2019}.

\section*{Acknowledgements}

BKG and LVEK acknowledge the financial support from a NOVA cross-network grant. BKG acknowledges the National Science Foundation grant AST-1836019. FGM, MK and AS acknowledge support from a SKA-NL roadmap grant from the Dutch Ministry of OCW. This work was supported in part by ERC grant 247295 ``AARTFAAC" to RAMJW. LOFAR, the Low Frequency Array designed and constructed by ASTRON, has facilities in several countries, that are owned by various parties (each with their own funding sources), and that are collectively operated by the International LOFAR Telescope (ILT) foundation under a joint scientific policy. This research made use of publicly available software developed for LOFAR and AARTFAAC telescopes. Here is a list of software packages used in the analysis: \textsc{aartfaac2ms} (\url{https://github.com/aroffringa/aartfaac2ms}), \textsc{aoflagger} (\url{https://gitlab.com/aroffringa/aoflagger}), \textsc{dppp} (\url{https://github.com/lofar-astron/DP3}), \textsc{wsclean} (\url{https://gitlab.com/aroffringa/wsclean}) and GPR foreground removal code (\url{https://gitlab.com/flomertens/ps_eor}). The analysis also relies on the Python programming language (\url{https://www.python.org}) and several publicly available python software modules: \textsc{astropy} (\url{https://www.astropy.
org}; \citealt{astropy2013}), \textsc{gpy} (\url{https://github.com/SheffieldML/GPy}), \textsc{emcee} (\url{https://emcee.readthedocs.io/en/stable/}; \citealt{emcee2013}), \textsc{matplotlib} (\url{https://matplotlib.org/}), \textsc{scipy} (\url{https://www.scipy.org/}) and \textsc{numpy} (\url{https://numpy.org/}).

\section*{Data availability}
The data underlying this article will be shared on reasonable request to the corresponding author.

\bibliographystyle{mnras}
\bibliography{ACE.bib} 


\bsp	
\label{lastpage}
\end{document}